\documentclass{emulateapj}
\shorttitle{ Multiplanet System HR 8799 }
\shortauthors{ Fabrycky and Murray-Clay }
\usepackage{natbib}

\newcommand{\mi}{M_{\rm in}}
\newcommand{\mo}{M_{\rm out}}
\newcommand{\ai}{a_{\rm in}}
\newcommand{\aout}{a_{\rm out}}

\begin{document}

\slugcomment{Accepted to ApJ -- Dec. 25, 2009}

\title{ Stability of the directly imaged multiplanet system HR 8799: resonance and masses }

\author{Daniel C.~Fabrycky\altaffilmark{1} and Ruth A. Murray-Clay\altaffilmark{2} }
\affil{Harvard-Smithsonian Center for Astrophysics\\
	60 Garden St, MS-51, Cambridge, MA 02138}
\email{daniel.fabrycky@gmail.com}

\altaffiltext{1}{Michelson Fellow}
\altaffiltext{2}{ITC Fellow}

\begin{abstract}

A new era of directly imaged extrasolar planets has produced a three-planet system \citep{2008M}, where the masses of the planets have been estimated by untested cooling models.  We point out that the nominal circular, face-on orbits of the planets lead to a dynamical instability in $\sim$$10^5$~yr, a factor of at least $100$ shorter than the estimated age of the star.  Reduced planetary masses produce stability only for unreasonably small planets ($\lesssim 2$~$M_{\rm Jup}$).  Relaxing the face-on assumption, but still requiring circular orbits while fitting the observed positions, makes the instability time even shorter.  A promising solution is that the inner two planets have a 2:1 commensurability between their periods, and they avoid close encounters with each other through this resonance.  That the inner resonance has lasted until now, in spite of the perturbations of the outer planet, leads to a limit $\lesssim 10 M_{\rm Jup}$ on the masses unless the outer two planets are \emph{also} engaged in a 2:1 mean-motion resonance.  In a double resonance, which is consistent with the current data, the system could survive until now even if the planets have masses of $\sim20$~$M_{\rm Jup}$.  Apsidal alignment can further enhance the stability of a mean-motion resonant system.  A completely different dynamical configuration, with large eccentricities and large mutual inclinations among the planets, is possible but finely tuned.
\end{abstract}


\section{Introduction} \label{sec:intro}

The method of direct imaging for the discovery of extrasolar planets has yielded spectacular first results over the last several years \citep{2004C, 2008L, 2008M, 2008K, 2009Lagrange}.  Direct imaging is a method for discovering planets located far from their host stars, an as-yet unexplored region of parameter space, and it promises new opportunities to characterize the planets using their own radiation.  However, because the gravitational influence of directly-imaged planets is not measured and the astrometric orbital arcs obtained so far are short, determining the planetary masses and orbital architectures of these systems is challenging.

In the newly-discovered planetary system HR 8799 ($=$ HD 218396), three planets have been imaged at projected separations of 24, 38, and 68 AU from their host star \citep{2008M}.  The best current estimate of their masses is derived from the planetary luminosities, measured in the infrared.  Because these planets are young and massive, they are still radiating prodigiously as they contract, cool, and become more gravitationally bound.  The masses are estimated using untested models of this contraction and cooling process.  One class of such models, the ``hot-start'' models, provides the largest luminosity possible at a certain mass and age, given assumptions about opacities in the planetary atmosphere.  Hot-start models have initially extended envelopes and a large entropy per baryon; even hotter models converge to a common track after a few Myr \citep{2002B}.  Therefore, for a given age and luminosity, these models should provide a lower limit on the mass.  For HR 8799, the lower-limit masses are $5$-$11$, $7$-$13$, and $7$-$13$~$M_{\rm Jup}$ for planets b, c, and d, respectively, based on a rather uncertain stellar age of $30$-$160$~Myr\footnote{ This estimate is given by \cite{2008M} based on four age indicators: Galactic space motion, main-sequence fitting, stellar pulsations, and the massive debris disk.  The first is circumstantial but consistent with the quoted ages; the others suggest an age $\lesssim 100$~Myr.  That the star has reached the main sequence suggests that it is at least several $10$s of Myr old. }, which is presumably also roughly the age of the planets.  

The following simple calculation illustrates why a lower mass limit can be inferred from a planet's contraction luminosity.  For HR 8799, the planetary luminosities have been measured to be $L \simeq 10^{-5} L_\odot$, and radii of $R \simeq 1.2 R_{\rm Jup}$ were derived from the objects' temperatures, measured by fitting photometry with a variety of synthetic spectral energy distributions \citep{2008M}.  Because the objects are cooling, they were more luminous in the past, so they have radiated at least $L t_{\rm age} \gtrsim 4 \times 10^{43}$~erg.  Their current binding energy, which supplied this luminosity, is $\simeq G M^2 R^{-1} \simeq 3 \times 10^{43}(M/M_{\rm Jup})^2$~erg, where the radius is roughly independent of the mass for Jupiter-mass objects.  Consequently, $M > 1 M_{\rm Jup}$.  Cooling models also take into account that $L$ diminishes with time, and thus arrive at a considerably larger mass.  Whether this larger calculated mass is a robust lower limit depends on the accuracy of the model.  Recently, \cite{2009D} measured the dynamical masses for a system of brown dwarfs (both of mass $\approx 57 M_{\rm Jup}$) and showed that cooling models overpredict the component masses by $\sim$25\%. 

If energy is lost during the process of planet formation, then an even larger planet mass would be needed to generate the currently observed luminosity.  For example, in the planetary core-accretion models of \cite{2007M}, considerable luminosity is radiated in the accretion stream and shock, and that energy is not internalized by the planet.  At the end of formation the planet has less gravitational potential energy to later supply its luminosity.  The integrated luminosity since formation would not account for the planet's current binding energy, so the mass needed to supply an observed luminosity at a given age may be much bigger.

The HR 8799 system has survived an order of magnitude longer than the primordial gas disk, which, if typical of disks of A stars, lasted $\lesssim3$~Myr \citep{1993H,2005H}.  The system has therefore had time to dynamically evolve in the absence of gas.
Though the planets orbiting HR 8799 are separated by tens of AU, the inferred minimum masses of the planets were large enough that their mutual gravitational interactions are important.  For example, a planet with mass $M_p = 10$~$M_{\rm Jup}$ orbiting a star of mass $M_* = 1.5$~$M_\odot$ at semi-major axis $a = 40$~AU dominates gravitational dynamics within its Hill radius of size $R_H = a(M_p/3M_*)^{1/3} = 5$~AU.  Because $R_H$ is a large fraction of the planetary separation, gravitational interactions among the planets can substantially modify the dynamical evolution of the system.

\begin{figure}
\epsscale{1.1}
\plotone{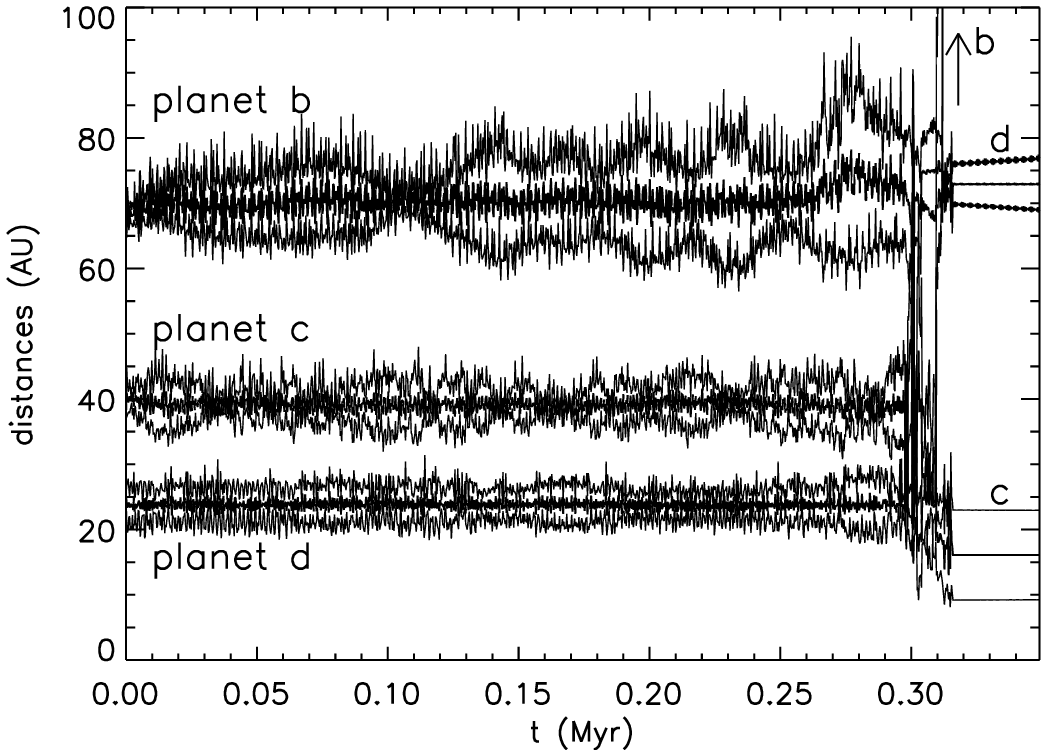}
\caption{ Semi-major axis, periapse, and apoapse for the three planets as a function of time in a numerical integration of the nominal model of the system, which has circular, face-on orbits, and planetary masses $M_b = 7 M_{\rm Jup}$, $M_c = 10 M_{\rm Jup}$, and $M_d = 10 M_{\rm Jup}$.  An instability occurs only $3\times10^5$~yrs into the integration, in which planets b and c suffer a close encounter, suggesting the planetary masses or orbits of the nominal model are in error. 
\vspace{0.1 in}}
\label{fig:ae}
\end{figure}

In fact, the nominal orbits reported in the discovery paper \citep{2008M} are unstable.  We integrated the Newtonian equations of motion of the proposed system using the Bulirsh-Stoer (BS) algorithm of the {\slshape Mercury} \citep{1999C} package (version 6.2), with an accuracy parameter of $10^{-12}$.  The planets are assigned circular, face-on orbits, and we used the nominal masses for all four bodies: $7$, $10$, $10$~$M_{\rm Jup}$ for planets b, c, and d, respectively, and $1.5$~$M_\odot$ for the star\footnote{For the specific initial conditions of this and other integrations herein, see Tables~\ref{tab:logI} and \ref{tab:logII}.}.  Figure~\ref{fig:ae} shows the results for the semi-major axis and maximum radial excursion of each planet as a function of time.  A close encounter between planets c and d at $0.298$~Myr (i.e., they enter within one Hill radius of one another) leads to a brief interval of strong scattering which ejects planet b at $0.316$~Myr (i.e., it reaches $>500$~AU with positive energy, and is removed from the simulation).  Planets c and d swap orbits and finish in a stable configuration, with no further semi-major axis evolution, but they exhibit a regular secular eccentricity cycle with a period of $1.5$~Myr.  This evolution is not unique in its details since the orbital evolution is chaotic.  However, qualitatively similar evolutions are common for simulated planetary systems constructed to match the discovery data: instability usually sets in well before the star's age of $\gtrsim 30$~Myr.  

The goal of this paper is to determine orbits that are consistent with the astrometric data, the inferred planetary masses, \emph{and} with dynamical stability over the system's age.  Neglecting stability considerations, there is a large amount of freedom in fitting orbits to the discovery data, because (1) the measured astrometric arcs cover only $\sim$2\% of the middle orbit and $\sim$1\% of the outer orbit, (2) the velocity of the inner planet is almost entirely unconstrained, and (3) the line-of-sight positions and velocities of the planets relative to the star are unknown.  {\it A priori}, two classes of orbital architectures are possible---those in which the planets occupy roughly coplanar orbits and those with large mutual inclinations.  Since planets form in disks, it is likely that they initially occupy nearly coplanar orbits, and systems that remain stable indefinitely are likely to stay roughly coplanar.  Alternatively, the system may not be indefinitely stable.
While old compared to the lifetime of the protoplanetary disk, the current age of the planetary system is probably less than one tenth the main-sequence lifetime of the star ($\sim$$1.5$~Gyr; \citealt{1967I}).  Without further analysis, it is thus possible that the planets are in the process of scattering off of one another, currently have large eccentricities and mutual inclinations, and will not be stable over the lifetime of the star.  In fact, current models predict that planetary systems undergo periods of strong mutual excitation, perhaps generically leading to the ejection of planets (e.g., \citealt{lld98}, \citealt{gls04}, \citealt{2009SM}, \citealt{2009VCF}).

\begin{deluxetable*}{lcccc}
\tablecaption{Astrometric constraints\label{tab:nom}}
\tabletypesize{\normalsize}
\tablewidth{0pt}

\tablehead{
\colhead{Planet} &
\colhead{[$x_E$, $x_N$]  (AU)} &
\colhead{$s$ (AU)} &
\colhead{[$v_E$, $v_N$]  (10$^{-3}$ AU day$^{-1}$)} &
\colhead{$v_p$  (10$^{-3}$ AU day$^{-1}$)}
}

\startdata
b	&[$60.16(6), 31.50(6)$] & $67.91(6)$&	[$1.49(0.14),  -2.18(0.14)$] & $ 2.64(0.14)$\\
c	&[$-25.90(6), 27.76(6)$] &$37.97(6)$& [$2.15(0.14), 2.47(0.14)$]& $ 3.27(0.14)$\\
d	&[$-8.45(6), -22.93(6)$] &$ 24.44(6)$& [$-3.5(2.7),  0.0(2.7)$] & ---
\enddata
\vspace{0.1 in}
\tablecomments{   \scriptsize{Sky-projected positions and velocities of each planet relative to the star, found by a rectilinear least-squares fit to the astrometry of Table 1 of \cite{2008M}.  Positions are at the epoch of 2008 Aug. 12, velocities assume no detectable orbital acceleration.}
\vspace{0.2 in}
}

\end{deluxetable*}

To explore these possibilities systematically, we take the following approach. We start with restrictive assumptions about the orbital architecture of the system, and we then progressively relax those assumptions.  At each stage, we find parameters that maximize the stability, and we finally argue that a resonant configuration is most likely for the system to have survived to its current age.  In \S\S\ref{sec:astrom}--\ref{sec:planarei}, we assume that the orbits of the planets are close to coplanar.  In \S\ref{sec:astrom} we discuss astrometric constraints on the orbits and show that, although the data are consistent with circular and coplanar orbits, the system orientations that fit the data best do not generate stable orbits.  In \S\ref{sec:masses} we determine what planetary masses would be needed for circular, coplanar orbits to be stable and argue that they are too low given the observed luminosities.  Having thus ruled out circular, coplanar solutions, we next allow the planetary eccentricities to vary.  Since the inner planet's eccentricity is unconstrained by the data, we first scan over non-circular orbits for the inner planet while keeping the outer two planets on circular orbits (\S\ref{sec:stable}).  The suggestive results of this experiment led us to our preferred configuration for the planetary system: a mean motion resonance between the inner two planets (\S\ref{sec:mmr}).  An initial exploration of the parameter space of possible resonant orbits shows that if the outer two planets are also in a mean motion resonance, the system could be stable even if the companion masses are twice as large as the nominal masses.  In \S\ref{sec:montecarlo}, we allow all the orbital parameters to vary via a Monte-Carlo method. We confirm that mean-motion resonance is the most likely reason the planetary system has survived.  We also find a scattering-type configuration that is stable for $30$~Myr, but we argue that it is unlikely.  In \S\ref{sec:conclude} we discuss our conclusions.

\section{ Astrometric Constraints } \label{sec:astrom}

\begin{figure}[b]
\epsscale{1.1}
\plotone{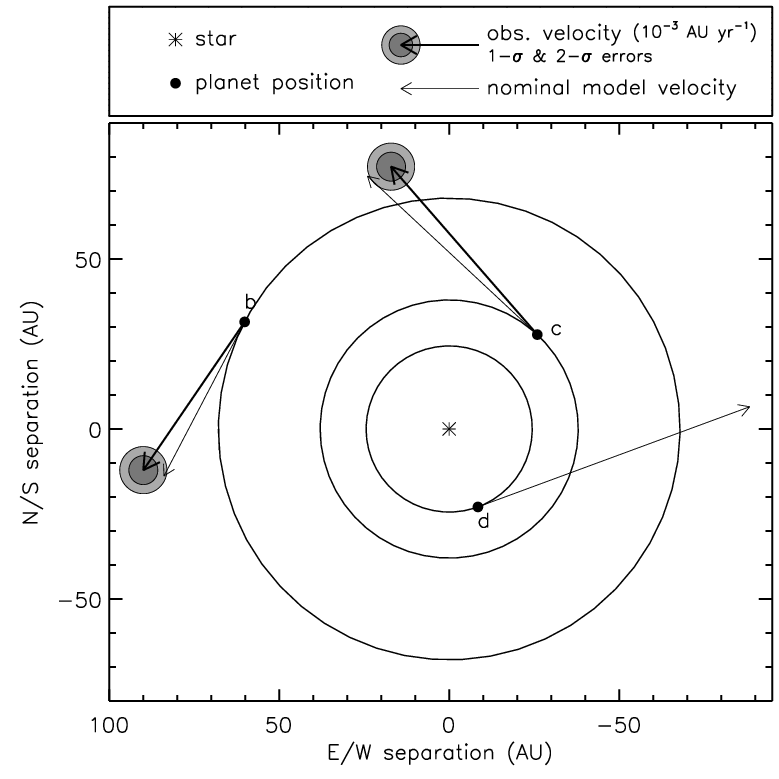}
\caption{ Observed sky-projected positions and velocities for the three planets, along with the velocities of face-on, circular orbits for $M_\star=1.5 M_\odot$ (the nominal model; model A of Tables~\ref{tab:orient} and \ref{tab:logII}).  The circles represent 1-$\sigma$ and 2-$\sigma$ errors on the measured velocities.  The inner planet, d, has a barely-detected velocity due to a short time baseline. Errors on the positions lie within the circles marking the locations of the planets. 
\vspace{0.1 in}}
\label{fig:data}
\end{figure}

In Figure~\ref{fig:data}, we plot the sky-projected position and velocity vectors (Table \ref{tab:nom}) of the three planets, at the epoch 2008 Aug. 12, as determined by least-squares fit to the astrometry in Table 1 of \cite{2008M}.  The distance to the star is $39.4 \pm 1.1$~pc, based on the {\slshape Hipparchos} parallax \citep{2007VL}.  We use this nominal distance to convert observed angular separations to AU.  The 3\% error thus introduced into the distances and velocities does not change our qualitative conclusions; in \S\ref{sec:montecarlo} we take this error into account.

\begin{deluxetable*}{cccccccccc}
\tablecaption{Solutions of circular, coplanar systems\label{tab:orient}}
\tabletypesize{\normalsize}
\tablewidth{0pt}

\tablehead{
\colhead{Solution} &
\colhead{$M_\star$ ($M_\odot$)} &
\colhead{$i$} &
\colhead{$\Omega$} &
\colhead{$\chi^2$} &
\colhead{d.o.f.} &
\colhead{$\alpha$} &
\colhead{$a_b (AU), \lambda_b$} &
\colhead{$a_c (AU), \lambda_c$} &
\colhead{$a_d (AU), \lambda_d$} 
}

\startdata
A & $\equiv 1.5$ & $\equiv 0^\circ$ & $\equiv 0^\circ$ & $12.63$ & $6$ & $0.049$ & $67.91$, $  62.36^\circ$ & $37.97$, $ 316.99^\circ$ & $24.44$, $ 200.23^\circ$ \\
B & $1.44 $ & $\equiv 0^\circ$ & $\equiv 0^\circ$ & $12.19$ & $5$ & $0.032$ & $67.91$, $  62.36^\circ$ & $37.97$, $ 316.99^\circ$ & $24.44$, $ 200.23^\circ$ \\
C & $\equiv 1.5$ & $21.3^\circ$ & $151.5^\circ$ & $9.07$ & $4$ & $0.059$ & $ 72.89$, $  62.30^\circ$ & $38.15$, $ 315.97^\circ$ & $ 25.47$, $ 202.23^\circ$ \\
D & $1.86$ & $33.2^\circ$ & $145.9^\circ$ & $5.67$ & $4$ & $0.225$ & $ 81.04$, $  61.30^\circ$ & $38.16$, $ 315.29^\circ$ & $ 27.69$, $ 204.92^\circ$ \\
E & $2.28$ & $41.4^\circ$ & $143.6^\circ$ & $2.77$ & $3$ & $0.429$ & $ 90.02$, $  60.20^\circ$ & $38.16$, $ 314.81^\circ$ & $ 30.33$, $ 207.30^\circ$
\enddata
\vspace{0.1 in}
\tablecomments{ \scriptsize{ The symbol ``$\equiv$'' denotes values among the parameters $M_\star$, $i$, $\Omega$ that are held fixed for this solution.  The inclination $i$ is the angle between the planetary system's orbital angular momentum vector and the vector towards the observer, and the ascending node $\Omega$ is measured East of North (so the position angle of a planet as it passes through the plane of the sky, toward the observer, is $\Omega$).  $\alpha$ is the significance value of $\chi^2$ being this high using a $\chi^2$-test, given a certain number of degrees of freedom (d.o.f.), under the null hypothesis that the circular coplanar model with the given orientation and stellar mass is correct.  For instance, the probability of observing these velocities given the nominal model is $<5\%$.} \vspace{0.2 in}
}
\end{deluxetable*}

The impression given by Figure~\ref{fig:data} is that we are seeing the planetary system face-on, with counter-clockwise, nearly circular orbits.  This is what we call the ``nominal model,'' and we plot the implied orbits and velocity vectors for a $1.5$~$M_{\odot}$ star, also in Figure~\ref{fig:data}.  
If all of the orbits are truly face-on and circular, their sky-projected separation $s\equiv\sqrt{x_{\rm E}^2+x_{\rm N}^2}=a$, and their sky-projected velocity $v_p\equiv \sqrt{v_{\rm E}^2 + v_{\rm N}^2} = v_{\rm orb}$, the orbital velocity.  
Since all of the planets are bound mostly by the mass of the star, they should follow circular orbits at semi-major axis $a$ with velocities $v_{\rm orb} = 2 \pi$~AU~yr$^{-1}$ $(M_\star/M_\odot)^{1/2} (a/\rm{AU})^{-1/2}$.  
For the outer two planets, $s$ and $v_p$ are measured with high precision (Table \ref{tab:nom}), providing two independent measurements of the stellar mass.  Given this nominal model, the stellar mass binding planet b is $M_{\star b}=1.60 \pm 0.17 M_\odot$ and the stellar mass binding planet c is $M_{\star c}=1.38 \pm 0.12 M_\odot$.  These values bracket the value of $M_{\star}=1.47 \pm 0.30 M_\odot$ preferred by combining parallax, magnitude, and spectroscopic information \citep{1999GK} and are in reasonable agreement: $\Delta M_{\star} \equiv M_{\star b}-M_{\star c} = 0.22 \pm 0.21 M_{\odot}$.  However, there is some tension in the observed velocities.  For both planets b and c, the observed velocity vector is $\sim$$2 \sigma$ away from perpendicular to the separation vector (from the star to the planet).  
The instability reported in the introduction is, however, the main failing of the nominal model.  

To address this failing, we first search for another model in which the planets are still coplanar and circular, but the system plane is inclined by an angle $i$ to the plane of the sky, with an ascending node $\Omega$ measured East of North, and a to-be-determined consistent mass $M_\star$.  The sky projection changes the magnitudes and directions of the velocity vectors and the inferred spacings of the planets, and taking it into account could lead us to infer a wider-spaced, more stable system.  We focus only on circular and coplanar models in this section, saving more complicated direct fits to the data for \S\ref{sec:montecarlo}.  The velocity field on the sky due to this model is:
\begin{equation}
\left( \begin{array}{c} v_E \\ v_N \end{array} \right) = n(x_E, x_N) \left( \begin{array}{c} - \alpha \sin \Omega \cos i -\beta \cos \Omega (\cos i)^{-1} \\ \alpha \cos \Omega \cos i -\beta \sin \Omega (\cos i)^{-1} \end{array} \right),
\end{equation}
where
\begin{equation}
\left( \begin{array}{c} \alpha \\ \beta \end{array} \right)= \left( \begin{array}{cc} \cos \Omega & \sin \Omega \\ -\sin \Omega & \cos \Omega \end{array} \right) \left( \begin{array}{c} x_E \\ x_N \end{array} \right),
\end{equation}
and 
\begin{equation} 
n(x_E, x_N) = (G M_\star)^{1/2} (\alpha^2 + \beta^2 (\cos i)^{-2})^{-3/4}
\end{equation}
is the mean motion as a function\footnote{Here we neglect the few-percent contribution of the planetary mass} of position.

We solve for the three parameters $i$, $\Omega$, and $M_\star$, assuming that the planets are on non-interacting Keplerian orbits, each of which only feels the mass of the central star.  We calculate $\chi^2$ values using $v_E$ and $v_N$, and their associated measurement errors, for all three planets (Table~\ref{tab:nom}; 6 data points)---we neglect the errors on $x_E$ and $x_N$, which are too small to affect our results. 

\begin{figure}
\epsscale{0.9}
\plotone{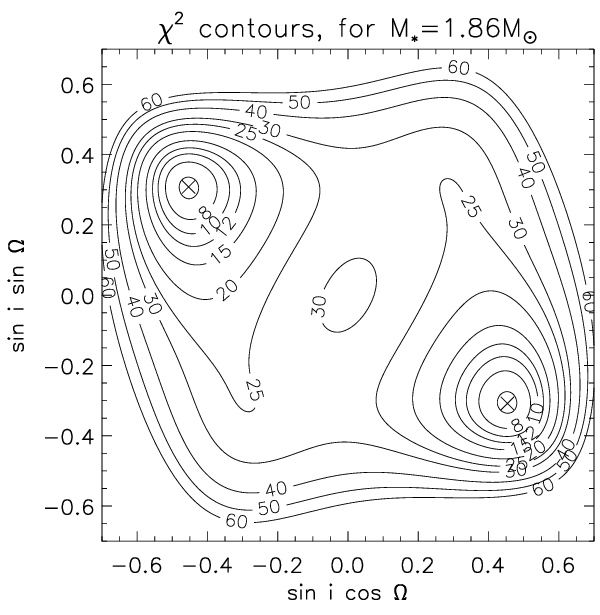}
\caption{ Model $\chi^2$ as a function of orbit orientation of circular, coplanar models at stellar mass of $M_\star=1.86 M_\odot$.  Note the degeneracy $\Omega \rightarrow \Omega + 180^\circ$, which arises because we are modeling an unobserved $z$ and $v_z$ (the direction $\hat{z}$ is away from the observer) for each planet, but due to the sky-projection inherent in the observations, these values could just as as well be $-z$ and $-v_z$, switching the ascending and descending nodes at $z=0$.  A face-on orientation lies at the origin.
\vspace{0.1 in}}
\label{fig:xisqcontour}
\end{figure}

\begin{figure}
\epsscale{0.9}
\plotone{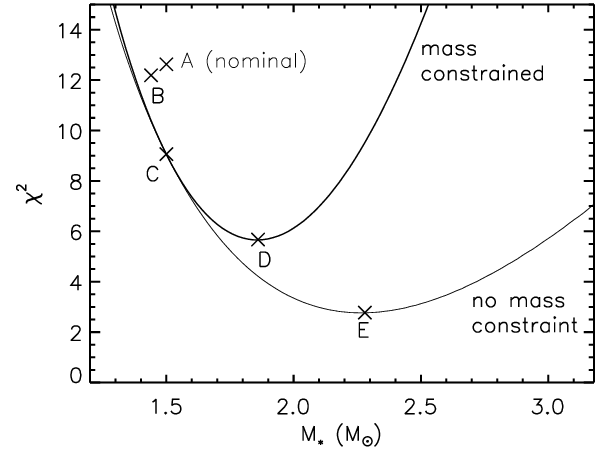}
\caption{ Model $\chi^2$ as a function of stellar mass, minimizing over system orientation ($i$ and $\Omega$) of circular, coplanar models.  Masses above the nominal mass $1.5 M_\odot$ are preferred.  The different crosses, each assigned a letter, are different solutions as given in Table~\ref{tab:orient}.
\vspace{0.1 in}}
\label{fig:mass}
\end{figure}

\begin{figure}
\epsscale{1.1}
\plotone{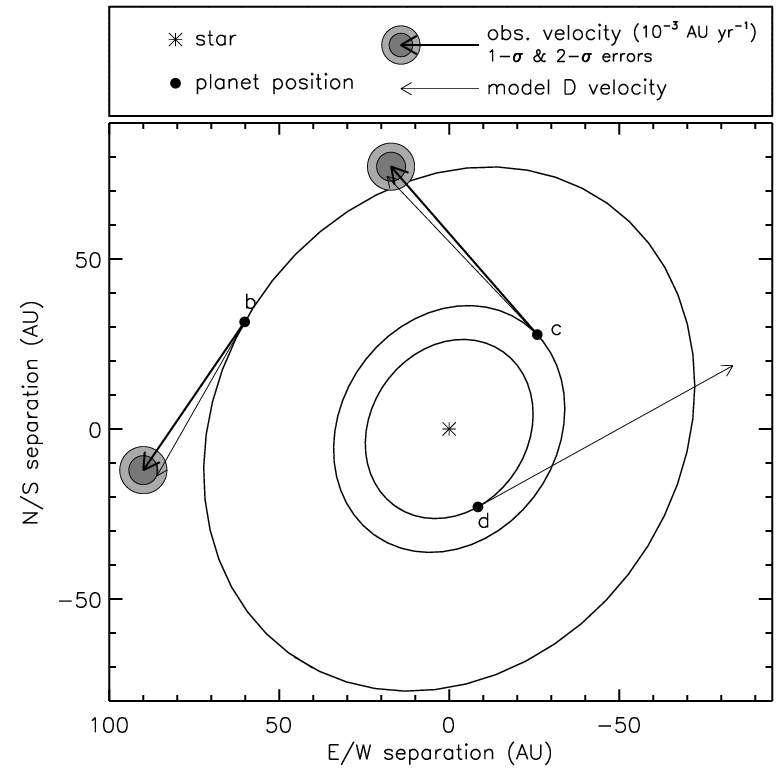}
\caption{ Observed sky-projected positions and velocities for the three planets, along with the velocity predictions of model D (Tables~\ref{tab:orient} and \ref{tab:logII}) of a circular, coplanar system.
\vspace{0.1 in}}
\label{fig:185}
\end{figure}

Solutions are reported in Table~\ref{tab:orient}.  In model A, which is the nominal model, we fix the parameters to their nominal values to serve as a baseline.  In model B, we require face-on orbits, but let $M_\star$ float, the result being not far from the nominal stellar mass.  In model C, we fix $M_\star=1.5 M_\odot$, but let the orientation float.  The orbits depart from face-on by $\sim$$20^\circ$, and $\chi^2$ improves a little.  In model D, we let all three parameters float, but respect the independently measured stellar mass by including $[(M_\star/M_{\odot}-1.5)/0.3]^2$ in $\chi^2$.  In model E, all three parameters float with no such mass constraint.  The orientation-dependence of $\chi^2$ is shown in Figure~\ref{fig:xisqcontour}, and the mass-dependence is shown in Figure~\ref{fig:mass}.  Interestingly, the best fits are for $M_\star$ much larger than the nominal value $1.5 \pm 0.3 M_{\odot}$.  This is not surprising given the good agreement of circular orbits because we are introducing line-of-sight offsets and velocities, so a more massive star is needed to make such orbits circular.  Figure~\ref{fig:185} shows how the velocity vectors of model D falls into the $1$-$\sigma$ error ellipse for each planet.  However, the inner two orbits are closer spaced than the nominal model, and the instability is even more rapid: in an integration a close encounter occurred between c and d on their second conjunction (see Table~\ref{tab:logI} for initial conditions).  

This integration and all those in \S\S\ref{sec:astrom}--\ref{sec:stable} were performed using the HYBRID integrator of {\slshape Mercury} with a timestep of 100 days.
Each integration was terminated when any two planets passed within 1 Hill radius of each other, one was ejected (distance to the star $>500$~AU with positive energy), or the system lasted $160$~Myr.  Before the onset of close encounters, energy was conserved to 1 part in $\sim$$10^5$ and angular momentum was conserved to 1 part in $\sim$$10^{12}$.  Though we used the HYBRID integrator, because the integrations were halted at the first close encounter, the integrator's treatment of close encounters did not affect our results.  In \S\ref{sec:ejection}, we verify that after a close encounter, at least one planet would be quickly ejected.

Similarly, we fit the best orientation for $M_\star$ values between $1.1$-$3.0 M_\odot$, spaced by $0.01 M_\odot$, and integrated those orbits.  No three-planet systems generated in this way were stable for more than $1.5\times10^5$~yr.  Therefore we find that more careful fits to the data, under the hypothesis of circular coplanar orbits, do not simply lead to a stable solution.

\section{ Much Lower Planetary Masses? } \label{sec:masses} 

Before relaxing the assumption that the planets' orbits are circular and coplanar, we ask how low the planets' masses must be for the nominal orbits to be stable.  Intuitively, if their masses are very small, the planets will not significantly perturb each other on the timescale of $30-160$~Myr.  
There is a well-developed framework for quantifying long-term stability in systems with only two planets.  In three-body systems, conservation of total energy and angular momentum constrains the possible motions \citep{1982MB}.  Applied to a system of a star and two planets, we may define Hill stability as a constraint that the planet that is initially closer to the star stays closer to the star for all time.  When the criterion for Hill stability is satisfied, a close encounter between the planets is prohibited (although escape of the outer planet to infinity, or the collision of the inner planet with the star, is not forbidden).  
Qualitatively, stability requires that the planets be separated by more than a few
mutual Hill radii: $R_{\rm H} \equiv \onehalf (\ai+\aout) \epsilon$, with $\epsilon \equiv [(\mi + \mo) / (3 M_\star)]^{1/3}$.  Define $\Delta$ as the planets' difference in semi-major axes in terms of $R_{\rm H}$. \cite{1993G} gave the Hill stability criterion as:
\begin{equation}
\Delta > \Delta_{crit} \equiv 2 \sqrt{3} [ 1 + 3^{1/2} \epsilon - \left( \frac{11 \mi+7 \mo}{18 M_\star} \right) 3^{-2/3} \epsilon^{-2} + ...] . \label{eq:hillbound}
\end{equation}
Evaluating these numbers using the nominal system with nominal masses $7$, $10$, $10$~$M_{\rm Jup}$, we have $\Delta_{cd}=2.68$ and $\Delta_{{\rm crit}, cd}=4.03$ for the inner two, and $\Delta_{bc}=3.69$ and $\Delta_{{\rm crit}, bc}=3.98$ for the outer two.  Apparently \emph{both} sub-systems fail to satisfy the Hill stability criterion.  

\begin{figure}
\epsscale{1.1}
\plotone{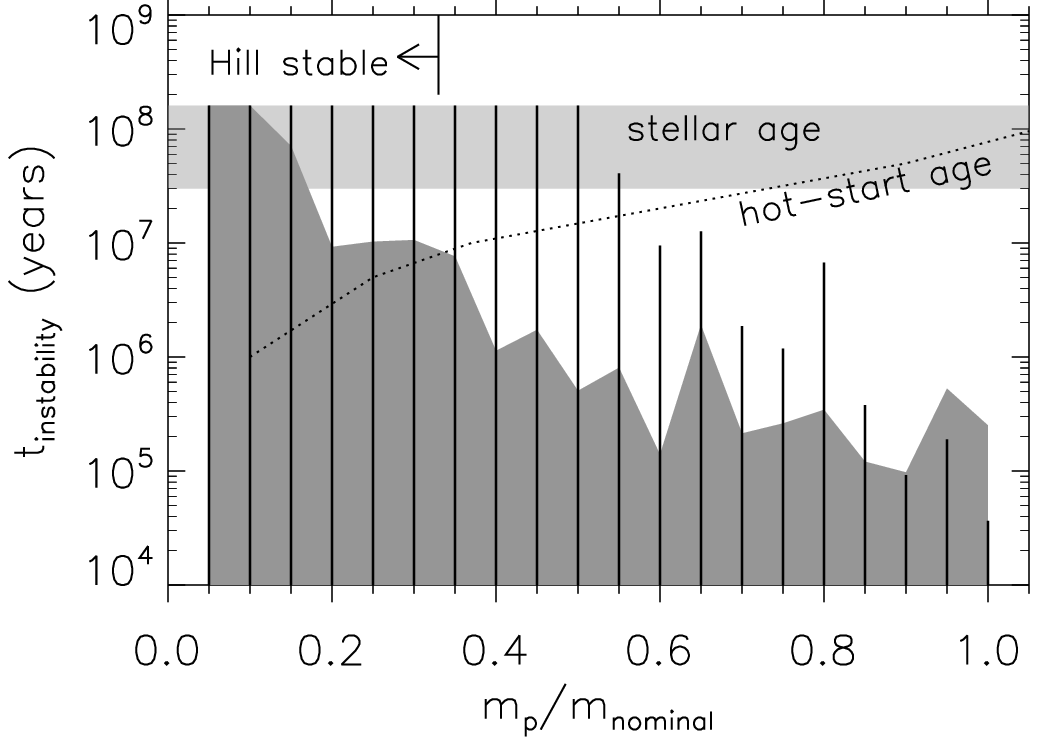}
\caption{ Time to instability versus scaling of planetary masses.  In each simulation, the planets are given circular, coplanar orbits (model A), but all of their masses are scaled down by a common factor from their nominal values (see Table~\ref{tab:logII}).  \emph{Solid vertical lines}: results for the inner two planets, in the absence of planet b.  \emph{Dark gray region}: results for all three planets.   \emph{Light gray region}: the stellar age as given by \cite{2008M}.  \emph{Dashed line}: time at which hot-start models of planets from \cite{2003Baraffe} reach luminosities of $10^{-4.7} L_\odot$ for masses scaled to $M_{\rm nominal} = 10 M_{\rm Jup}$.  Given the system age estimate from \citet{2008M}, stability requires planet masses $\lesssim 2 M_{\rm Jup}$.  Cooling models allow for stability of planets up to $\sim$$3.5 M_{\rm Jup}$ at the cost of an uncomfortably low system age: $\sim$10 Myr.
\vspace{0.1 in}}
\label{fig:massmult}
\end{figure}

We performed numerical simulations to find just how small the planets would need to be to remain stable.  We are helped by the long orbital periods and short system age (only $\sim$$10^6$ dynamical times), which allows suites of integrations to be rather inexpensive.  First we surveyed the instability near the nominal orbits (``A''), as the search of \S\ref{sec:astrom} did not reveal any more stable starting points.  Let us focus on the inner sub-system (c-d), as it is further from stability, and ask the question: by what factor must we multiply the nominal masses for stability over $30$~Myr?  In Figure~\ref{fig:massmult} we plot the time to instability --- when the first Hill-sphere entry occurs --- versus this common mass scaling.  Vertical lines represent two-planet systems consisting of planets c and d on circular, face-on orbits.  We note that below $M_p = 0.33 M_{\rm nominal}$, where Hill's stability criterion is satisfied, all of the two-planet systems last for $160$~Myr, when the integrations were stopped.  \citet{1993G} found that if the planets initially have small eccentricity (radial excursions comparable or less than a Hill radius) and are not in resonance, then the timescale for instability drops rapidly after this boundary (eq.~[\ref{eq:hillbound}]) is crossed.  However, instability does not appear on timescales relevant for the c-d subsystem until $M_p \simeq 0.5 M_{\rm nominal}$.  In a separate suite of integrations (not plotted), we found that the nominal orbits and masses of the outer pair of planets can be stable for 160~Myr.  We also plot the instability timescale of the three-planet system (dark gray region), with each of the three planetary masses scaled by a common factor.  The masses must be lower than about $1/5$ of the nominal masses to remain stable $30$~Myr, the lower limit on the stellar age (depicted by the light gray stripe).  

When considering three or more planets, there are no sharp stability boundaries, but there are well-established empirical scaling relations between semi-major axis separation and instability timescale \citep{1996C,2007Z}.  Applying the scaling relation of \citet[appendix A, fit 1]{2008Chatterjee} to the HR 8799 system implies $\Delta \gtrsim 4.4$ if the system is to remain stable $\gtrsim 30$~Myr.  Let us assume the instability between the inner two planets is dominant, so this limit applies to $\Delta_{cd}$; then the masses must be $\lesssim 1/4$ of the nominal masses, in good agreement with the non-resonant systems of Figure~\ref{fig:massmult}.  We note, however, that none of the published scaling relations extend to planetary-to-stellar mass ratios of $5\times10^{-3}$, nor do they strictly apply if adjacent planets are unequally spaced in Hill radii ($\Delta_{cd} \neq \Delta_{bc}$), both of which are relevant for HR 8799.  For masses $>1/4$ of the nominal masses, our results give a longer lifetime than the scaling of \citet{2008Chatterjee}, sometimes orders of magnitude longer.  Regardless, we confirm that instability can occur even if the sub-systems are initially Hill stable.  In circular, face-on orbits, the implied upper limits of masses --- $1.5$, $2$, and $2$~$M_{\rm Jup}$ --- are incompatible with any cooling model at ages greater than $30$~Myr, even extreme hot-start models.

\section{ Non-circular inner orbit? } \label{sec:stable}

In the previous section, we found that face-on, circular orbits, consistent with the astrometric constraints, could only be stable if the planetary masses were implausibly low.  In this section, we choose the lowest planetary masses that are compatible with hot-start models, and we choose a non-circular orbit for planet d (its orbit is currently unconstrained by observations).  We seek systems that remain stable until the lower limit on the stellar age of $30$~Myr.

\begin{figure}
\epsscale{1.1}
\plotone{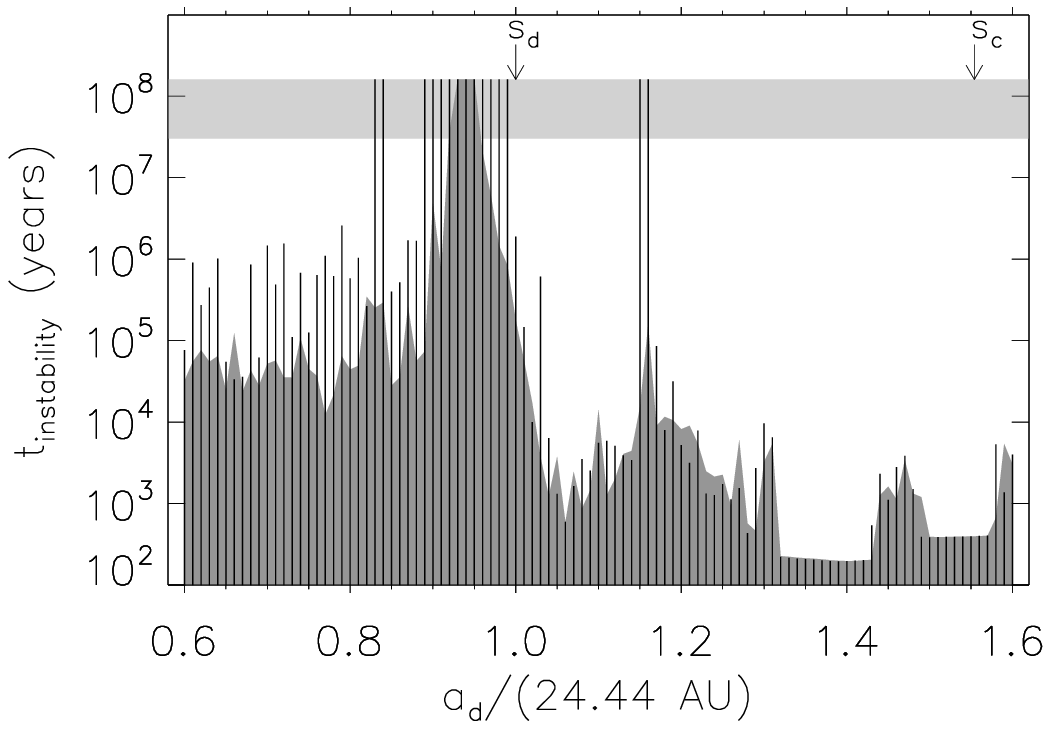}
\caption{ Instability time of coplanar systems, with planets b and c on circular face-on orbits and planet d on a non-circular orbit, with masses of $5$, $7$, and $7$~$M_{\rm Jup}$ for b, c, and d respectively (see Table~\ref{tab:logII}).  \emph{Lines}: based on planets c and d (in the absence of planet b) from the nominal model, but choosing a non-zero eccentricity for planet d to satisfy its currently-observed distance from the star.  \emph{Dark gray region}: same as before, but in the presence of planet b with its nominal parameters.  \emph{Light gray region}: the stellar age as given by \cite{2008M}.  A few three-planet systems last the stellar age, and these correspond to a 2:1 mean motion resonance between planets c and d.  Current separations between the star and the planets are labeled $s_d$ and $s_c$. 
\vspace{0.1 in}}
\label{fig:instabgridelm}
\end{figure}

We first simulate the inner two planets, each of $7 M_{\rm Jup}$, in the absence of planet b.  They are given coplanar orbits, with planet c on a circular orbit at $a_c=s_c =37.97$~AU.   The initial longitudinal separation is given by the observed positions, assuming face-on orbits ($\lambda_c - \lambda_d \approx 117^\circ$).   We scan over a grid of semi-major axes for the inner planet.  For $a_d<s_d = 24.44$~AU, $e_d$ is chosen so that apastron is at $24.44$~AU, and for $a_d>24.44$~AU, $e_d$ is chosen so that periastron is at $24.44$~AU (see Table \ref{tab:logII} for how the initial conditions are generated).  These choices maximize the chance that the two-planet system will be stable, while matching the constraint of the currently-observed separations from the star.  We plot the instability times in Figure~\ref{fig:instabgridelm} as vertical lines.  We repeat this calculation with planet b present with its nominal orbital elements (see Table~\ref{tab:logII}) and with mass $5$~$M_{\rm Jup}$, and plot those instability timescales in Figure~\ref{fig:instabgridelm} as a gray region.  

\begin{figure}
\epsscale{1.1}
\plotone{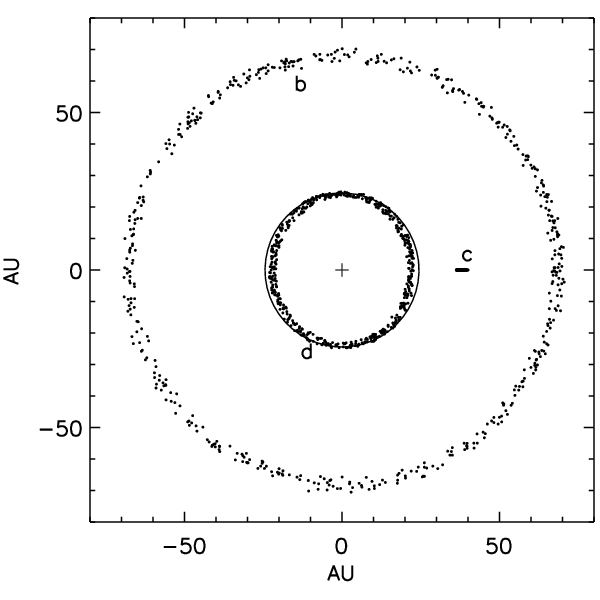}
\caption{ Positions of the planets, every $\sim3 \times 10^5$~yr, in the numerical integration with $a_d = 0.95 \times 24.44 {\rm AU} = 23.22 {\rm AU}$ from Fig.~\ref{fig:instabgridelm} (see also Table~\ref{tab:logII}).  The rotating coordinates are centered on the star with planet c on the positive horizontal axis.  The circle is a distance from the star of $24.44$~AU.  When the inner planet lags the middle planet by $\simeq 117^\circ$ (its current position), its distance from the star is $\simeq24.44$~AU, but when it reaches the middle planet's longitude, it is always closer to the star.  Planets are labeled near their currently-observed positions relative to one another.  The 2:1 mean motion resonance protects the two planets from close encounters and adds coherence to the long-term transfer of energy and angular momentum during encounters.
\vspace{0.1 in}}
\label{fig:protect}
\end{figure}

We observe that a very narrow range of $a_d$ is compatible with both the observed astrometry of the planets and with dynamical stability.  The presence of the third planet narrows this range still further.  The center of this range corresponds with the 2:1 mean motion resonance between planets c and d.  (The position is offset from the location $a_d = (1/2)^{2/3} a_c$ because the large mass ratios induce fast precession.)  In Figure~\ref{fig:protect} we show how this resonance protects the planets from close encounters.

We ran identical simulations with planetary masses of $7$, $10$, and $10~M_{\rm Jup}$, and found qualitatively similar results, except the most stable three-planet simulation lasted only $10$~Myr.  In the next section we examine this resonant protection mechanism and find initial conditions that produce acceptably long survival times even for these and even higher masses.

\section{Mean motion resonance} \label{sec:mmr}

Inspired by the fact that Figure~\ref{fig:instabgridelm} shows a region of greater stability in the vicinity of the 2:1 resonance between c and d, we search for a face-on system near the center of the resonance.  We use the BS integrator throughout this section.  We find a solution in the absence of planet b in which the resonance angle, 
\begin{equation}
\phi_d = 2\lambda_c - \lambda_d - \varpi_d, \label{eqn:phi}
\end{equation}
librates with small amplitude around $0^\circ$.  When evaluating resonant angles, we compute the orbital elements with astrocentric coordinates.  Resonance requires that $a_d$ is low enough for planet d's period to be commensurate with planet c's, and the observations require that $e_d$ is high enough for planet d to reach its currently observed separation from the star.  Currently, we observe $\lambda_c - \lambda_d \approx 117^\circ$, so $\phi_d\approx0^\circ$ implies $\lambda_d-\varpi_d\approx 126^\circ$.  Thus we find small libration is compatible with planet d being closer to apoastron than to periastron at the current time, in which case the velocity should be smaller than that of a circular orbit at the same distance.  Integration with these initial conditions for planets c and d, in the absence of b, indeed shows libration and long-term stability (at least 160 Myr), for initial $a_d$ and $e_d$ values that place the planets in resonance (e.g., the system labeled ``two-planet resonance'' in Table~\ref{tab:logI}).  

\begin{figure}
\epsscale{1.1}
\plotone{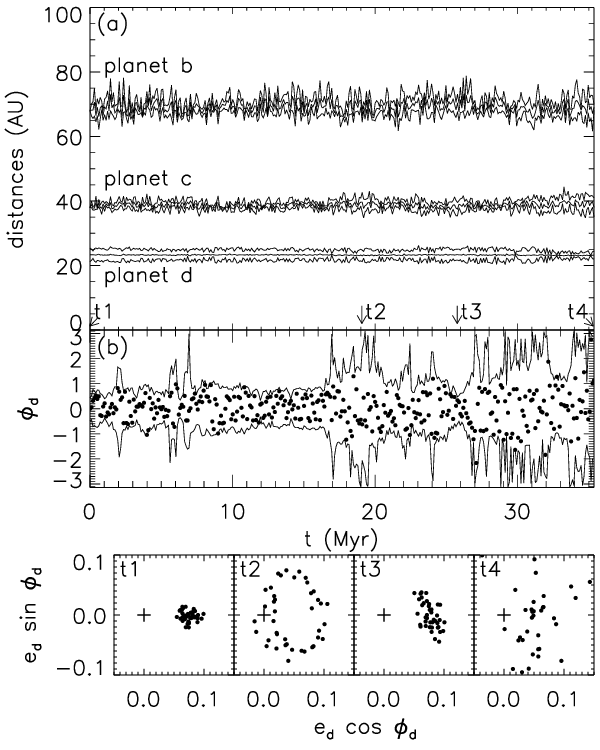}
\caption{ A simulation of the nominal masses, which is initially protected from close encounters by the 2:1 resonance between planets c and d, but it is destroyed after $35.6$~Myr due to interactions between planets b and c (see Table~\ref{tab:logI} for initial conditions).  (Panel a) Semi-major axis, periapse, and apoapse for the three planets as a function of time; (panel b) \emph{dots}: resonance angle every $\sim10^5$~yr (libration is rapid, on nearly orbital timescales, and is not well-sampled) and \emph{lines}: its running envelope, as a function of time; (panels t1-t4) phase plot of the resonance angle, over short durations, as labeled in panel b.
\vspace{0.1 in}}
\label{fig:unstabres}
\end{figure}

In such solutions, the resonance angle for planet c usually does not librate.  The resonance involves only the eccentricity of planet d.  When planet b is added, planets b and c excite each other's eccentricities and cause the libration amplitude of $\phi_d$ to fluctuate.  Sometimes these excited eccentricities cause an encounter between b and c; sometimes the loss of libration in $\phi_d$ allows an encounter between c and d.  In Figure~\ref{fig:unstabres} we show an example of this instability, where planets b and c start in their nominal orbits, $a_d=23.32$~AU, $e_d=0.09$, $\phi_d = 0^\circ$, and all bodies have their nominal masses (see Table~\ref{tab:logI}).  Panel (a) shows the range of motion of each orbit versus time, panel (b) shows the resonance angle versus time, and the bottom panels show brief segments ($3\times 10^4$~yr, at times labeled above panel b) of the motion of the resonance angle through phase space.  Over such brief intervals, the libration amplitude holds rather steady, except at the very end of the integration.  In this example, the instability causes an encounter between planets c and d at $35.6$~Myr.

\begin{figure}
\epsscale{1.1}
\plotone{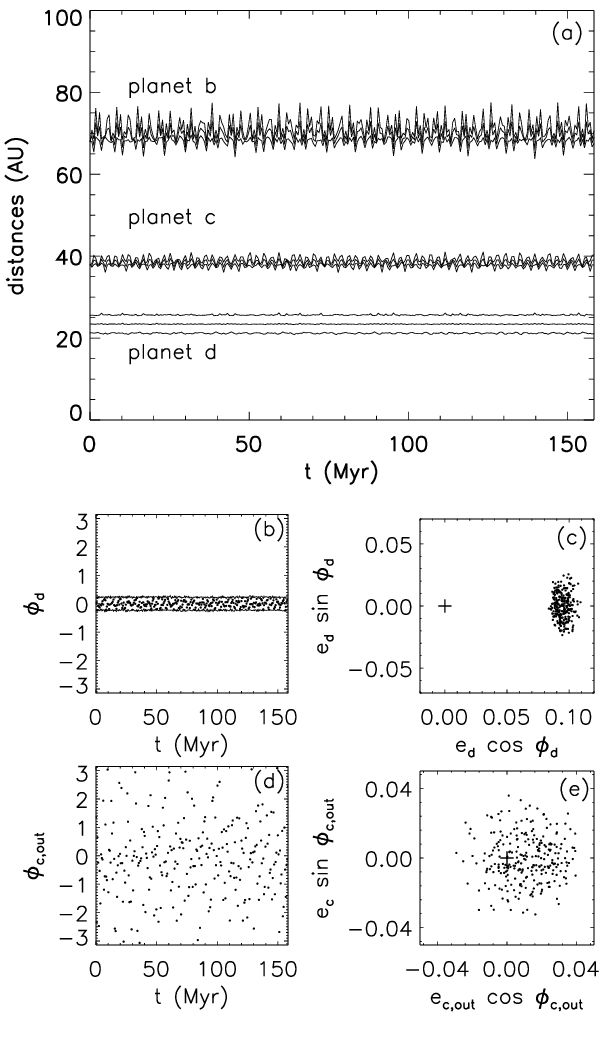}
\caption{ A simulation of the nominal masses, lasting for 160 Myr with no signs of imminent instability, due to a double resonance (see Table~\ref{tab:logI} for initial conditions).  Panels are as in Figure~\ref{fig:unstabres}.  The resonance angles are defined as $\phi_d = 2\lambda_c - \lambda_d - \varpi_d$ and $\phi_{c,out} = 2\lambda_b-\lambda_c-\varpi_c$.
\vspace{0.1 in}}
\label{fig:boundres}
\end{figure}

Compared to the non-resonant cases, this system showed considerable longevity: it lasts long enough to be a plausible model for the observed system.  We have found a way to calm the strongest interactions, those that cause instability after a few thousand orbits: a resonance between planets c and d that protects them from close encounters.  This resonance protects the system until the somewhat longer timescale interactions between b and c cause an instability.  But those interactions can also be suppressed by postulating yet another resonance.  We integrated the nominal masses with initial conditions as above except $a_d=23.42$~AU instead of $23.32$~AU (Table~\ref{tab:logI}).  The resulting system showed resonance protection between planets b and c.  The 2:1 resonance is active, which this is possible far from its nominal location because the pericenters are precessing on nearly orbital timescales.  In Figure~\ref{fig:boundres} we show this system lasting for 160 Myr. In this example, the resonance angle $\phi_d$ is librating with small amplitude the whole time (panels b and c), and the resonance angle $\phi_{c,out} = 2\lambda_b-\lambda_c-\varpi_c$ spends more time near $0^\circ$ (panels d and e), indicating the system is protected by both resonances.  Even after 160 Myr of evolution, we have verified that there are epochs at which this solution fits the astrometric data of Table~\ref{tab:nom}.  We found print-outs for which a rotation in the plane of the sky matched the simulated to the observed positions within a fractional error of 1\% (more print-outs would likely find a closer match), and then we calculated $\chi^2$ based on the velocities of Table~\ref{tab:nom}.  The resulting $\chi^2=11.4$ was both acceptable and quite competitive with the models of \S\ref{sec:astrom}.

\begin{figure}
\centering
\scalebox{1.1}
{\plotone{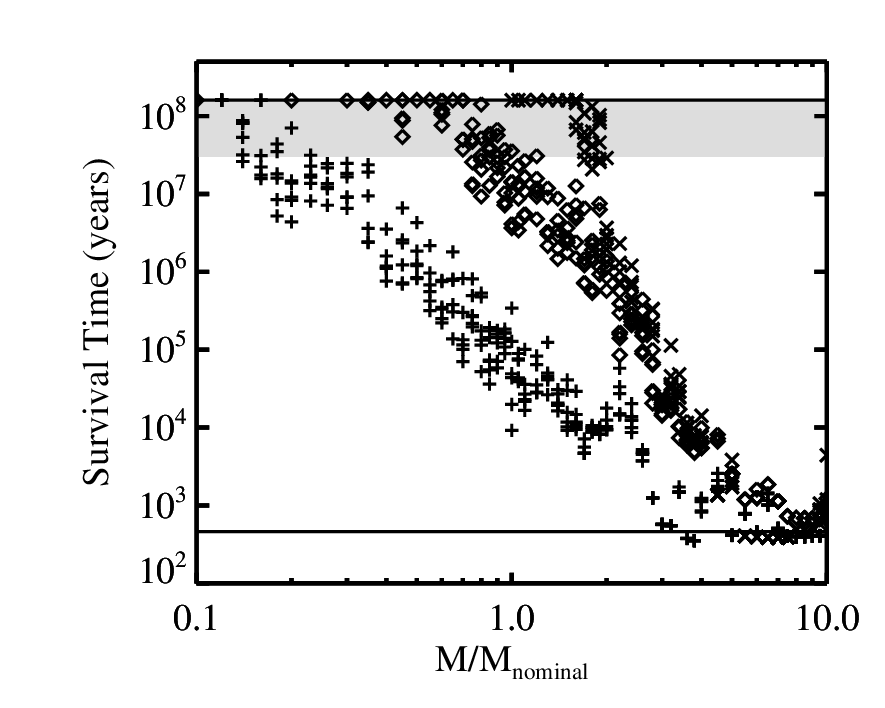}}
\caption{ Time to instability of three-planet systems as a function of a common mass scaling for the planets relative to their nominal masses.  See Table~\ref{tab:logI} for initial conditions.  To test the sensitivity of our results to small changes in initial conditions, for each mass scaling, we calculate the time to instability for the stated orbital parameters and for five additional sets of orbital parameters generated as follows.  For each orbital element, we draw a random number from a normal distribution with mean 0 and a standard deviation of 1.  We then multiply the result by a scaling factor and add it to the initial value of that element.  The scaling factors are $10^{-4}$ AU for the semi-major axis, $10^{-4}$ for the eccentricity, and $0.01^\circ$ for the inclination, ascending node, longitude of pericenter, and mean anomaly.  \emph{Plusses:} Face-on, circular orbits;  \emph{Diamonds:} Orbits in which the 2:1 resonance between planets c and d is active initially; \emph{Crosses:} Orbits in which both the 2:1 resonance between planets c and d and the 2:1 resonance between planets b and c are active initially; \emph{Gray region:} the stellar age as given by \cite{2008M}. }
\label{fig:massfactor}
\vspace{0.1 in}
\end{figure}

The next step is to understand how these resonances protect the system as a function of planetary mass.  For instance, Figure~\ref{fig:instabgridelm} shows four integrations in which the resonance allows planets of masses $M_b = 5$,  and $M_c = M_d = 7$~$M_{\rm Jup}$ to be stable for $30$~Myr, which is consistent with the observed system.  But can the system survive at the nominal masses with only one resonance?  How high can the masses go, in the double resonance?  In Figure~\ref{fig:massfactor}, we plot the time to instability for a wide range of planetary mass scalings.  We use initial conditions corresponding to the nominal face-on, circular orbits (non-resonant), the initial conditions for Figure~\ref{fig:unstabres} (singly resonant), and parameters chosen to maximize stability of the double resonance for massive planets.  All are listed in Table \ref{tab:logI}. Because the resonant locations shift with increasing planetary mass, the ideal orbital parameters for stable resonance depend on the masses.  In a suite of integrations we slightly vary the initial conditions (see Table~\ref{tab:logI}) to sample the chaotic outcomes.  

We find that systems with the nominal masses rarely survive 30 Myr with a single resonance, but can easily survive at least 160 Myr with a double resonance.  In fact, our integrations show that a doubly-resonant system can be stable for 160 Myr, even for planetary masses a factor of two larger than the nominal values.  That is, if this doubly-resonant configuration is correct, the planets could even have the masses of brown dwarfs.  We find it remarkable that a double 2:1 resonance can allow planetary masses an order of magnitude larger than the $\sim$$2 M_{\rm Jup}$ allowed in a stable, non-resonant system. 

In some of these integrations, we have found the three-body Laplace resonance, with angle $\phi_L = \lambda_d - 3 \lambda_c + 2 \lambda_b$, librating temporarily (see also \S\ref{sec:mode} below).  Such solutions are also consistent with the astrometric data.  The Laplace angle was first observed to librate in the satellites of Jupiter (e.g., \citealt{1999MD}).  Besides HR 8799, two other extrasolar planetary systems have been proposed to inhabit the Laplace resonance.  Extra peaks in the periodogram of radial velocity residuals of the GJ 876 system \citep{2005R} and the HD 82943 system \citep{2008Beau} could correspond to planets in the Laplace resonance with the known planets.  Although each of these three extrasolar systems taken separately is merely suggestive of 4:2:1 and Laplace resonances, taken together they are quite intriguing.  They may point to a new area of research in multiplanet systems that has been explored rather little so far, both theoretically and observationally.

\section{ Monte-Carlo search } \label{sec:montecarlo}

In the integrations so far, we have systematically varied a few parameters, concluding that a mean motion resonance is a promising solution to stabilize the system.  Now we seek alternatives by allowing all the other orbital parameters to vary.  The objective is to survey what orbits are allowed when the age of the system and the planetary masses are presumed to be robust.  To be conservative, we adopt the youngest system age ($30$~Myr), corresponding to the least massive planets $(5, 7, 7)$~$M_{\rm Jup}$, as in \S\ref{sec:stable}.  We select all the other variables with a Monte Carlo method.  We draw the stellar mass from a normal distribution with mean $1.47 M_\odot$ and standard deviation $0.30 M_\odot$ \citep{1999GK}, and we draw the system distance from a normal distribution with mean $39.4$~pc and standard deviation $1.1$~pc \citep{2007VL}.  The planetary sky-projected positions and velocities are drawn from normal distributions according to the observed parameters of Table~\ref{tab:nom}.  Note that these parameters are derived from the discovery observations of \cite{2008M} only; in \S\ref{sec:newdata} we check which systems are consistent with the important precovery observation of planet b by \cite{2009M}.  What remains is to draw $z$ and $v_z$ for each planet, which are not constrained by the observations; we make that choice in various ways in the following subsections.

In this section, we follow some planets with very high eccentricities and in some cases integrate through close approaches between planets.  We use the BS integrator as before, and we follow the integration until one planet is ejected or collides with the star.  Over 30~Myr, energy is typically conserved to better than one part in $10^6$, and angular momentum is typically conserved to better than one part in $10^7$.

\subsection{ From crossing orbits to ejection } \label{sec:ejection}

First, we wish to verify that once planets' orbits begin crossing, at least one of them is ejected in a timescale much shorter than the age of the system.  To do so, note that the expression for orbital energy of a single planet around a star is a monotonically increasing function of $|z|$ or $|v_z|$.  Therefore, selecting non-zero values for those parameters will lead to a planet that is less bound than in the case of a face-on orbit.  If $v_z=0$, there is a maximum value of $|z|$, called $|z|_{\rm max}$, that permits a bound orbit.  If $z$ is given, then there is a maximum value of $|v_z|$, called $|v_z|_{\rm max}$, that permits a bound orbit.  We wish to find how long it takes for planets on crossing orbits that are \emph{not} marginally bound to be ejected by each other, so we first selected $z$ from a distribution uniform in the interval $[-|z|_{\rm max}/3, |z|_{\rm max}/3]$, after which we selected $v_z$ from a distribution uniform in the interval $[-|v_z|_{\rm max}/3, |v_z|_{\rm max}/3]$.  This choice of distribution has the advantage of being connected to the observables (actually, complementary to them), being easy to implement, and being tuned to answer the question.  It has the disadvantage of not corresponding to a simple distribution in orbital element space; nevertheless, a very wide range of orbital elements are sampled. 

We ran 1530 systems generated in this way, integrating to an ejection of one component (defined as reaching an orbital distance $>500$~AU with positive energy).  The median time to ejection was $0.22$~Myr, and the maximum time was $7.7$~Myr.  These timescales are longer than the $\sim$$0.02$~Myr scattering phase of Figure~\ref{fig:ae} due to significant mutual inclinations.  In any case, the scattering phase will not contribute significant longevity to the system, and we are justified in stopping integrations at the first close approach in other sections of this paper.  \cite{2009VCF} have also reached this conclusion for the planets of HR 8799.

\subsection{ Moderately eccentric, coplanar planets } \label{sec:mode}

Next, we extend the analysis of \S\ref{sec:stable} to the case in which all three planets have non-zero eccentricities.  For simplicity, and acknowledging that the planets probably formed in a common, flattened disk, we first investigate coplanar systems.  There are several steps to generating a ($z$, $v_z)$ pair for each planet:
\begin{enumerate}
\item{draw stellar mass, distance, and planetary sky-projected positions and velocities, as described above;}
\item{draw a vector uniformly from the unit sphere, which serves as the direction of all the planets' angular momenta;}
\item{compute $z$ and $v_z$ for each planet, consistent with the already-chosen spatial variables;}
\item{ discard the system if $x$, a number drawn from a uniform distribution in $[0,1]$, is greater than $\mathcal{L}/\mathcal{L}_{\rm ML}$, where 
\begin{equation}
\mathcal{L} \equiv \prod_{j=b,c,d} e_j \exp [ - (e_j/\sigma)^2/2 ] \label{eq:ray}
\end{equation}
and
\begin{equation}
\mathcal{L}_{\rm ML} \equiv (\sigma \exp[-1/2] )^3
\end{equation}
with $\sigma = 0.05$. }
\end{enumerate}

If a system is discarded at steps 3-4, the process begins anew with step 1.  Step 4 is a technique called rejection sampling, and its purpose is to impose a prior distribution on the selected orbital elements $e_b$, $e_c$, and $e_d$.  We sought planetary orbits with low to moderate eccentricity, using the Rayleigh distribution (eq.~[\ref{eq:ray}]), as recommended by recent work on the generation of eccentricities by planetary perturbations \citep{2007Z,2008JT}.  We chose $\sigma$ consistent with the dynamically ``inactive'' population of \cite{2008JT}.  

We ran $16,581$ systems generated this way, until a close approach or $30$~Myr elapsed.  The median time until close approach was 3,100 yr, and only $49$ survived $30$~Myr.  Of the surviving systems, we checked for 2:1 resonances, two resonant arguments for the inner pair and two resonant arguments for the outer pair.  We considered the resonance to be dynamically significant if $h \equiv e \cos \phi$ had a non-zero average value (as in Figure~\ref{fig:boundres}, panel e): $|\langle h \rangle| > 2.5 \sqrt{\langle h^2 / n \rangle}$, where the averages were performed over $n$~($\sim$$100$) printouts of astrocentric orbital elements.  This criterion is considerably looser than traditional definitions of being ``in a resonance,'' either libration of a resonant angle or lying interior to a separatrix in phase space.  Nevertheless, this criterion indicates (i) protection against close encounters in the sense of Figure~\ref{fig:protect} and (ii) enhanced coherency to energy and angular momentum transfers during conjunction, as conjunctions occur at preferential phases of the orbit. All $49$ survivors had at least one of the four angles fulfilling this criterion: for $26$ only the inner pair were engaged in the resonance, for $2$ only the outer pair were engaged in the resonance, for $21$ both resonances were active.  We expect that most of these $49$ survivors will eventually be disrupted by the perturbations of planet b; in no case was the motion as periodic as in Figure~\ref{fig:boundres}.  Even systems stable for 160 Myr may become unstable over the main-sequence lifetime of the star \citep{2009GM}.  Although 21 integrations displayed the 4:2:1 double resonance, in only one case did the Laplace angle $\phi_L$ librate the entire time (about $180^\circ$), and that system shows the strongest inner and outer resonances of the entire set.  It is unclear if or how libration of $\phi_L$ enhances stability, over and above each pair of 2:1 resonances.

\begin{figure}
\epsscale{1.1}
\plotone{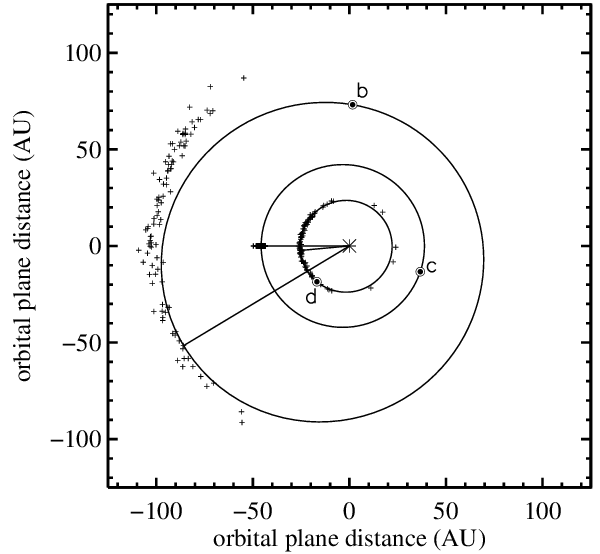}
\caption{ A simulation lasting for 109 Myr with weak mean motion resonance, but with strong apsidal locking (see Table~\ref{tab:logII} for initial conditions).  Small crosses correspond to the apoapse position of each planet, at snapshots spaced by $1.1$~Myr, in a frame centered on the star with planet c's apoapse on the negative horizontal axis.  The straight lines from the star are the current lines of apoapse, the ellipses are the current orbits of the planets, and the black dots with white rims are the current positions of the planets.  All three apsidal lines advance with the same average period of $1.5$~Myr.  The azimuthal distribution of the crosses indicates the variation in $\Delta \varpi$, the radial distribution of the crosses indicates the variation in eccentricity (until the very end, the semi-major axis change is negligible).  For each pair taken separately, the inner planet's apocenter tends to be co-located with the outer planet's apocenter, so close approaches are forbidden.  Note that at the current time planets b and c are near pericenter and planet d is near apocenter, enhancing their dynamically-packed appearance.
\vspace{0.1 in}}
\label{fig:seclock}
\end{figure}

One surviving system showed only a very weak inner 2:1 mean motion resonance.  We examined the first 1.6~Myr of this integration in detail, finding that $P_c/P_d$ fluctuated in the range $2.30-2.45$ and $P_b/P_c$ fluctuated in the range $2.7-3.0$.  With thousands of print-outs we determined that $h$ associated with $\phi_d$ had a non-zero average of $+0.012$.  However, its large range $-0.078$ to $+0.105$ suggests weaker protection by the 2:1 resonance than that enjoyed by the other stable systems.  The outer two planets occupied the 3:1 mean-motion resonance associated with the angle $\phi_{3:1} = 3\lambda_b-\lambda_c-2\varpi_c$ at a similarly weak level.  Most strikingly, all three planets of this system maintained apsidal alignment with one another throughout the integration: their relative apsidal angle $\Delta \varpi$ librated around $0^\circ$.  Apsidal locking apparently provides additional protection against close approaches: an inner planet only comes to apocenter at the same spatial location where an outer planet is at apocenter, so the two bodies do not come too close together.  This system\footnote{Initial conditions are given in Table~\ref{tab:logII}.} is illustrated in Figure~\ref{fig:seclock}, which gives a pictorial representation of the apsidal protection mechanism.  We also searched all of the stable systems for a tendency towards apsidal alignment, as quantified by $|\langle \cos \Delta \varpi \rangle|$ being non-zero in the same way as $h$ above.  We found that apsidal alignment was also common for systems with strong mean-motion resonance; in the set of 49 survivors it occurred 3 times between the inner planets only, 12 times between the outer planets only, and 10 times among all three planets. 

\subsection{ Arbitrarily eccentric, coplanar planets } \label{sec:planarei}

We repeated the procedure of \S\ref{sec:mode} without step 4, imposing no prior on the eccentricities.  However, to ensure that close encounters were not already happening, we only accepted each system if $a_d (1+e_d) < 0.85 a_c (1-e_c)$ and $a_c (1+e_c) < 0.85 a_b (1-e_b)$.  We integrated 25,280 systems, of which only 5 lasted 30 Myr, and the median time to a close approach was 37,000~yr.  Of the 5 survivors, one had the 2:1 mean-motion resonance active and secular alignment between the inner two planets.  In the other 4, all three planets tended towards alignment, and two of these had the 2:1  mean-motion resonance active between the outer two planets.  In 3 cases, the inner planet had $e_d>0.95$ and a current position near apocenter, and the apsidal alignment between d and c was rather tight: $|\varpi_c - \varpi_d| \lesssim 45^\circ$.  These systems may correspond to the non-linear secular resonance identified by \cite{2004MM}.  We consider it unlikely that configurations with strong apsidal alignment but no mean-motion resonance correspond to the true system, as explained below (\S\ref{sec:newdata}).

If all three planets orbit in the same plane, one may wonder whether the debris disk and the stellar equator share it as well.  From the observed positions and velocities, we can use this Monte Carlo study to constrain the orientation of the planets.  Coplanar systems that fit the observed positions and velocities with arbitrary eccentricities, but non-crossing orbits, have line-of-sight inclinations less than $45^\circ$.  The subset of stable systems obey this same limit.  Note that this is not substantially different from the limit for circular orbits, cf. Figure~\ref{fig:xisqcontour}.  Constraints on the stellar spin orientation---an expected rotational velocity $v$ and a measured $v \sin i$---led \cite{2009R} to derive a stellar inclination of $13^\circ-30^\circ$, consistent with this limit.

\subsection{ Circular, non-coplanar orbits } \label{sec:nonplanar}

Next, we investigate whether moderately non-coplanar orbits are substantially more stable, even in the absence of any resonance.  If not, we expect the conclusion that a resonance is needed applies to any roughly coplanar systems, not just strictly coplanar ones.  The following series of integrations is not intended to match the currently observed positions, but to be a parametric study of mutual inclination.  Relative to the nominal case (A), but now with lower-limit masses, we varied the initial inclination and node of planets b and d (see initial conditions in Table \ref{tab:logI}).  Both planets b and d are given the same inclination (relative to c, which always starts at $i_c=0$).  However, they are given different nodes, to sample the same inclination 36 times, so that the spread of chaotic outcomes are represented.  The systems were integrated until a close approach or 160 Myr.  

\begin{figure}
\epsscale{1.1}
\plotone{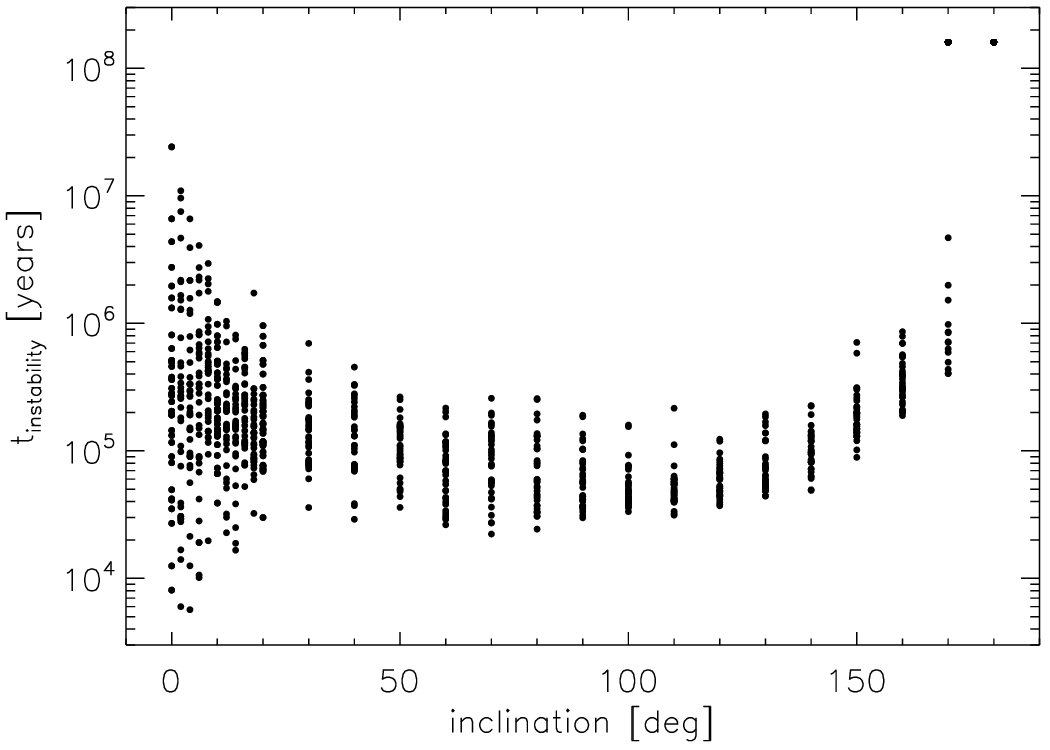}
\caption{ Inclination dependence of instability times.  Planet b and planet d are both initially inclined relative to planet c by the indicated inclination, and a $6 \times 6$ grid of initial conditions for $\Omega_b$ and $\Omega_d$ in [$0^\circ$, $60^\circ$, ... , $300^\circ$] was performed---see Table~\ref{tab:logI}. Increased mutual inclinations do not increase stability times for prograde orbits.
\vspace{0.1 in}}
\label{fig:mutinc}
\end{figure}

In Figure~\ref{fig:mutinc} we plot the resulting times of instability.  At low inclination the spread of these times is several orders of magnitude.  Varying the initial orientation angles, which also control the initial longitude of each planet, causes some systems to have a close approach within a few tens of orbits and causes other systems to last $>1$~Myr due to the protection afforded by the 2:1 resonance between planets d and c.  As the initial inclination increases, one might expect Hill-sphere encounters will be delayed, as the motion out of the plane exceeds a Hill radius at $i \simeq 6^\circ$ for these masses.  Nevertheless, we observe that the median instability time modestly decreases until $40^\circ$, is constant between $50^\circ$ and $140^\circ$, and increases dramatically from $150^\circ$ to $180^\circ$, perfectly retrograde.  The shorter instability timescale for substantially non-planar systems is likely due to the \cite{1962K} effect, which causes inclination to decrease and eccentricity to increase on a secular timescale of $\sim$$10^5$~years.

We conclude that we can likely extend our conclusions from strictly coplanar systems to systems with large prograde inclinations.  Systems with adjacent planets orbiting in the opposite sense can avoid instability; i.e., retrograde systems are inherently more stable than prograde systems, as has been shown in other contexts \citep{2003N,2008G}.  No mean-motion resonance or apsidal locking appears to be active in protecting retrograde systems from instability, but the brief timescales of conjunction may be responsible. We do not expect a retrograde configuration for planet d, nor is it consistent with the data (see \S\ref{sec:newdata}).

\subsection{ Arbitrary orbits } \label{sec:arbitrary}

In this subsection, we remove all restrictions on individual eccentricity and orbital orientation, to have a completely data-driven sampling of orbits.  We expect that not many systems generated this way are realistic, as planet d's orbit is so poorly constrained.  Nevertheless, after drawing the stellar and observable planetary parameters, we drew each $z$ uniformly from the interval $[-|z|_{\rm max}, |z|_{\rm max}]$ and then drew each $v_z$ uniformly from the interval $[-|v_z|_{\rm max}, |v_z|_{\rm max}]$.  We rejected the resulting system if either:
\begin{enumerate}
\item{ any of the planets initially had positive energy (despite trying to avoid this case by construction of the osculating orbital elements); or}
\item{ planetary orbits crossed, i.e., $a_d (1+e_d) > a_c (1-e_c)$ or $a_c (1+e_c) > a_b (1-e_b)$.}
\end{enumerate}

We integrated 3010 systems until the first ejection (not stopping at close approaches), or until $30$~Myr elapsed.  Of these, 13 systems survived the entire time.  Three cases had inner apsidal alignment, and another two cases had outer apsidal alignment.  A common characteristic is that planet d has a moderate-to-large eccentricity and comes to apocenter at its currently-observed position.  Thus the period ratio $P_c / P_d$ can be quite large, and the system rather hierarchical.  In fact, we found that using the full formulation of Hill stability \citep{1982MB}, the Star-d-c sub-system (neglecting planet b) was usually Hill stable if planet d is prograde with respect to the others.  With only small perturbations from planet b, the system remains stable for $30$~Myr, with the inner subsystem fulfilling Hill's stability criterion for most of that time. In contrast, as mentioned above, no particular protection mechanism was apparent for the retrograde systems.  The greater stability of retrograde orbits is not seen in Hill's stability calculations, just as the greater stability of retrograde satellites is not reflected in the usual Jacobi constant in the circular restricted three-body problem \citep{1997H}.

\begin{figure}
\epsscale{1.2}
\plotone{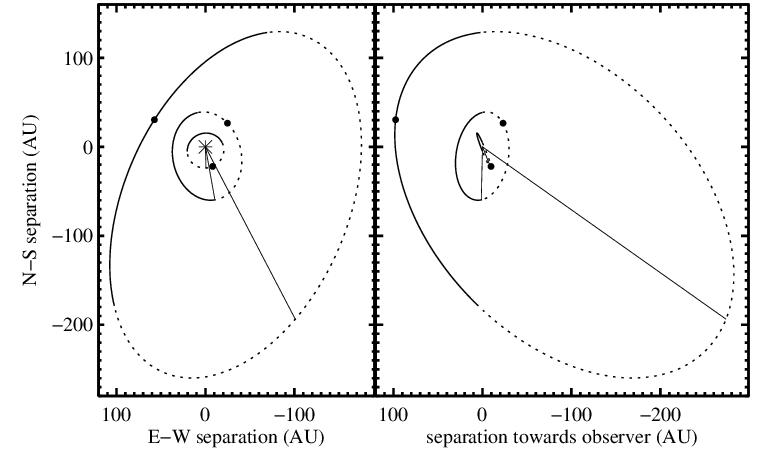}
\caption{ A non-coplanar system lasting 110 Myr before a close approach, shown at the current epoch.  See Table~\ref{tab:logII} for initial conditions.  The front view (left panel, observer's prospective) and right-side view (right panel) are given.  The parts of the orbit on the observer's side of the sky plane are solid, parts on the far side of the sky plane are dashed.  The apocenter lines are drawn, showing that planet d is currently near apocenter, and planets c and d are currently near pericenter, which is usually required of models that do not rely on protection from a mean-motion resonance.
\vspace{0.1 in}}
\label{fig:noncoplanar}
\end{figure}

A common attribute is that in \emph{all} 13 stable systems, planet b is in a very wide orbit, but it comes to pericenter at its currently-observed location.  One way to quantify this is to note that, in all cases, the mean anomaly of $b$ at the present epoch is within $10^\circ$ of $0^\circ$, which would happen only $1/18$ of the time for randomly phased orbits.  In such an orbit, it perturbs the inner two planets minimally, contributing to the system stability.  The orbits of one of these 13 systems at the current epoch is displayed in Figure~\ref{fig:noncoplanar}.

We regard these systems, with very large mutual inclinations and eccentricities, to be a possible, but not plausible, explanation of the observational data.  The outer orbit being very near periastron seems finely tuned, because it does not spend much time there.  It is as if it is swooping in from several hundred AU, just in time to have its picture taken by \cite{2008M}.  We also note that orbits like those of Figure~\ref{fig:noncoplanar} are currently consistent with circular, coplanar orbits, but they do not fulfill this condition at most other orbital phases.  Since the observed velocities are consistent with circular and coplanar, we can quantify the likelihood that the true system actually has very non-coplanar orbits, as follows.  We produced outputs for the system displayed in Figure~\ref{fig:noncoplanar}  at every $\sim3500$~years of a $10^8$~year integration, during which the semi-major axes and eccentricities showed no qualitative long-term changes.  For each of the outputs, we found the $\chi^2$ of the best-fitting circular, coplanar model, minimizing over $(M_\star, i, \Omega)$ as in \S\ref{sec:astrom}, with $M_\star$ in the range $[1,2]\times M_\odot$ with no penalty for being far from the nominal value.  We assign the observed error bars on velocities to the sky-projected velocities in the simulation, then compare them to the best-fitting velocities that come from a circular, coplanar hypothesis.  With the current snapshot (Figure~\ref{fig:noncoplanar}), the system fits a circular, coplanar model with $\chi^2=2.18$.  In only 639 of the 28,996 snapshots was $\chi^2$ better than this value.  So we conclude that the current phase of this particular non-coplanar system is fine-tuned to the $\sim$$2.2\%$ level.  We thus prefer models that actually \emph{are} close to circular and coplanar.

Another consideration disfavors these solutions with large mutual inclination and large eccentricities.  Fits to the infrared spectral energy distribution of HR 8799 suggest a population of colliding, dust-forming bodies with a semimajor axis of $\sim$$100$~AU, though this distance is still uncertain \citep{2006WA,srs+09,2009R}.  It is unlikely that planet b's orbit actually crosses this belt of bodies, which may constrain $e_b$ to less than a few tenths.  This constraint will be observationally accessible in the near future.

\subsection{ Comparison to New Data } \label{sec:newdata}

\begin{figure}
\epsscale{1.1}
\plotone{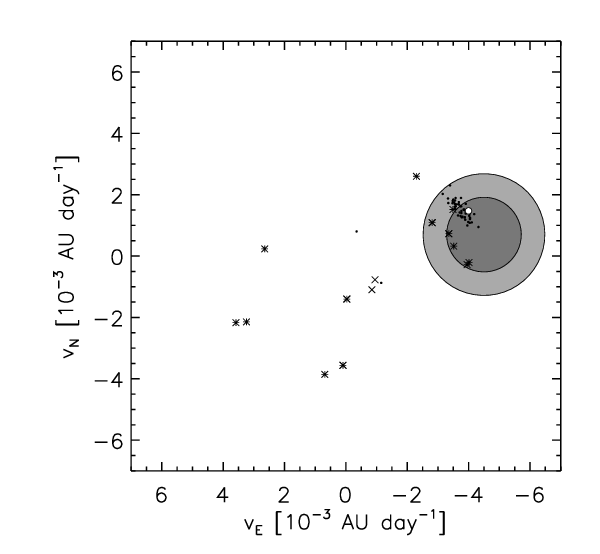}
\caption{ Sky-projected velocity of planet d from various stable systems that fit the discovery data \citep{2008M}, from \S\S\ref{sec:mode}, \ref{sec:planarei}, \ref{sec:arbitrary}.  The shaded regions are the 1-$\sigma$ (dark gray) and 2-$\sigma$ (light gray) regions allowed by the precovery observation by \cite{2009M}.  The open circle is the nominal (circular, face-on, $M_\star = 1.5 M_\odot$) solution.  Small dots are the solutions with 2:1 mean-motion resonance between either adjacent pair or a 4:2:1 double resonance.  Crosses are coplanar solutions for which no 2:1 mean-motion resonance was active, but apsidal alignment was preserved among the inner two planets.  Asterisks are solutions with arbitrary eccentricity and inclination from \S\ref{sec:arbitrary}.  
\vspace{0.1 in}}
\label{fig:predict}
\end{figure}

The foregoing analysis was based solely on the astrometric measurements reported in the discovery paper.  Between the original submission of this work and now, several new measurements were reported based on careful analysis of previously-collected images \citep{2009L,2009F,2009M}.  In particular, the measurement by \cite{2009M} of positions of planet d over a one-year baseline determines its sky-velocity to be 
\begin{equation}
[v_E, v_N] = [-4.5\pm0.8, 0.7 \pm 0.8]\times10^{-3} {\rm AU\,day}^{-1}.  \nonumber
\end{equation}
These values come from a regression of position versus time, in combination with the \cite{2008M} data, as in Table~\ref{tab:nom}.  For the purposes of this section, that value eliminates some of the previously possible dynamical configurations, as follows.  In Figure~\ref{fig:predict} we plot the 1-$\sigma$ and 2-$\sigma$ contours of the measured velocity of planet b, and overlay the orbits from the preceding sections.  We find that solutions with a retrograde planet d, and those for which planet d is at apastron of an very eccentric orbit, are now ruled out.  However, the 2:1 resonant solutions are still quite consistent with the new data, including the weakly-resonant, apsidally-locked system (Fig~\ref{fig:seclock}) and some of the very non-coplanar (yet non-resonant) solutions (e.g., Fig~\ref{fig:noncoplanar}).  Although these latter solutions fit the data, recall that we are seeing these systems at a special time, so we doubt they correspond to the true system.

\section{Discussion}  \label{sec:conclude}

We have investigated the orbital stability of the newly-imaged planetary system HR 8799.  The nominal orbital model and masses are not stable.  In fact, no model with circular, coplanar orbits that also fits the astrometry well is stable, regardless of the inclination and orientation of the system on the sky.  

To overcome this problem by reducing the planetary masses, values $\lesssim 2$~$M_{\rm Jup}$ are required. This can happen if the cooling models under-predict the luminosity, though that is difficult to understand, as even hot-start models cannot produce the observed luminosity at such low masses (see \S\ref{sec:intro}).  Such masses would be plausible if the system is considerably younger than expected, yet the star has reached the main sequence \citep{2008M}.

Our favored solution is that a 2:1 resonance between the inner two planets preserves stability.  Assuming that the inner pair of planets are in resonance, two qualitatively different configurations are possible:
\begin{itemize}
\item{The outer two planets are not in resonance. } This configuration remains stable in the perturbing presence of planet b only if the planetary masses are $\lesssim 10$~$M_{\rm Jup}$ (Fig.~\ref{fig:massfactor}). (It is also possible, given a less likely system orientation, that only the outer resonance is active.  This configuration leads to a similar mass limit.)
\item{The outer pair of planets are also in 2:1 resonance.  }
This solution fits all the current data for the system.  At the nominal masses, the system can easily survive for the age of the star.  In fact, the planetary masses could be up to $\sim1.9$ times bigger than their nominal values without violating stability constraints (Fig.~\ref{fig:massfactor}).  This value is similar to the maximum planetary masses that can stably exist at the fixed point of the 2:1 resonance \citep[fig. 8]{2003B}.  It will be very interesting to find a test of this hypothesis.  The Laplace angle can also librate in this system, though this is not a requirement for stability.
\end{itemize}
Strong apsidal alignment, while unnecessary for system stability, allows planetary survival in a weaker 2:1 mean-motion resonance.

One final type of system architecture is, in principle, consistent with stability of the HR 8799 system.  The planets may be hierarchically spaced, with large eccentricities and mutual inclinations, but inhabit phases of their orbits that look closely-packed at the moment.  Such systems fit the current data but are finely tuned both in their orbital parameters and in the time at which we are viewing the system.

It is possible that other stabilizing resonant configurations exist, but were missed because they occupy small regions of phase space.  The 2:1 mean-motion resonance dominates our randomly-generated stable systems and would naturally yield the observations without fine tuning, so we consider it to be the most likely stabilizing mechanism.

This study brings up several issues for future observations of HR 8799, and directly-imaged multiplanet systems in general, as follows.

	First, it serves as the first test of hot-start cooling models for exoplanets.  They barely pass the test if only planets d and c are in resonance, and they comfortably pass the test if all three planets are in resonance.  We hope more detailed dynamical studies of this system will sharpen this test as more data are collected.  If the doubly-resonant configuration can somehow be verified, it would considerably weaken this test.  We have not yet directly used dust observations as a dynamical constraint.  The spectral energy distribution reveals a massive debris disk surrounding the planetary system, with an orbital radius of $\gtrsim 66$~AU \citep{2006WA}.  Given that planet b is observed at $a_b \gtrsim 68$~AU, we expect that the inner edge of the debris disk must be $\gtrsim 90$~AU, and that future measurements and modeling will find that orbital radius to be plausible and even preferred.  Such a model could in turn serve as a complementary test of planet b's mass, in analogy to the test of the mass of Fomalhaut b \citep{2009C}.  Perhaps other directly-imaged systems will fortuitously arrange for complementary tests.
	
	Second, we found evidence of a mean motion resonance at very large orbital separations, much farther than those found by the radial velocity technique, of which there are many (e.g., \citealt{2001Marcy,2004Mayor}).  Sometimes resonant identification is based on stability arguments in those systems (e.g., \citealt{2005C}), as it is here.  The most commonly invoked evolutionary mechanism for trapping planets into resonance with one another is convergent migration in the protoplanetary disk.  The properties of migration in a disk with multiple massive planets deserve further investigation to determine the conditions under which convergent migration and resonance capture are possible at the locations of the HR 8799 planets.
We verified that if planets d and c were initially placed in circular orbits exterior to the 2:1 resonance, they are stable to collisions or ejections for $\gtrsim 30$~Myr, so getting into the resonance without first becoming unstable is not a problem in this case.  One difficulty with differential migration is that any additional migration, after the resonance is reached, efficiently increases the eccentricities.  That this requires implausible fine-tuning in the absence of eccentricity damping by the gas disk has been discussed for the 2:1-resonant GJ876 system \citep{2002LP}.
In a trial integration, we introduced a force to simulate outward migration of the inner planet, following \cite{2002LP}, with a timescale $a/(da/dt) = 10^7$~yr and no eccentricity damping, starting from double-2:1 resonance at the nominal masses (fig.~[\ref{fig:boundres}], table~\ref{tab:logI}).  The planetary eccentricities rapidly increased and the system began scattering after a semi-major axis change of $\sim15$\%, illustrating its fragility.  We expect calculations of migration into resonance will be very interesting for this system.  We also showed how perturbations by a third planet tend to disrupt a mean-motion resonance (when only the inner sub-system is in resonance).  This mechanism adds to a growing list of ways to disrupt resonances among planets, including turbulent fluctuations in a protoplanetary disk \citep{2008A}, tidal dissipation \citep{2007T}, and scattering of planetesimals \citep{2006M,2007Morby}.

	Third, it may seem surprising that dynamical stability arguments are needed to correctly solve the orbits of the first directly-imaged multiplanet system.  However, this situation is also common for multiplanet systems discovered by radial velocity.  For instance, \cite{2005V} and \cite{2006Lee} have found that dynamical stability can constrain orbital parameters more tightly than radial-velocity data alone.  Furthermore, planets that are discovered by direct imaging of their self-luminosity are biased to have high masses, making stability less assured.  The bias of direct imaging towards large angular separation implies that very long orbital periods will be common for such discoveries, so we foresee many stability analyses predicated on only the sky-projected positions and velocity vectors of planets.  We hope this paper proves to be a useful example of how to conduct such an analysis.

\acknowledgments
We thank Matthew Holman, Eric Ford, Jack Lissauer, Guillermo Torres, Sean Andrews, Yanqin Wu, Andrew Shannon, and members of CITA and U of Toronto Astrophysics department for helpful discussions.  We thank Scott Tremaine, Eugene Chiang, and the anonymous referee for suggesting that we expand the investigation to mutually inclined solutions.  DF is grateful for funding through a Michelson Fellowship, which is supported by the National Aeronautics and Space Administration and administered by the Michelson Science Center.  RM is grateful for an Institute for Theory and Computation Fellowship at Harvard University.

\bibliography{ms} \bibliographystyle{apj}

\begin{thebibliography}{52}
\expandafter\ifx\csname natexlab\endcsname\relax\def\natexlab#1{#1}\fi

\bibitem[{{Adams} {et~al.}(2008){Adams}, {Laughlin}, \& {Bloch}}]{2008A}
{Adams}, F.~C., {Laughlin}, G., \& {Bloch}, A.~M. 2008, \apj, 683, 1117

\bibitem[{{Baraffe} {et~al.}(2002){Baraffe}, {Chabrier}, {Allard}, \&
  {Hauschildt}}]{2002B}
{Baraffe}, I., {Chabrier}, G., {Allard}, F., \& {Hauschildt}, P.~H. 2002, \aap,
  382, 563

\bibitem[{{Baraffe} {et~al.}(2003){Baraffe}, {Chabrier}, {Barman}, {Allard}, \&
  {Hauschildt}}]{2003Baraffe}
{Baraffe}, I., {Chabrier}, G., {Barman}, T.~S., {Allard}, F., \& {Hauschildt},
  P.~H. 2003, \aap, 402, 701

\bibitem[{{Beaug{\'e}} {et~al.}(2003){Beaug{\'e}}, {Ferraz-Mello}, \&
  {Michtchenko}}]{2003B}
{Beaug{\'e}}, C., {Ferraz-Mello}, S., \& {Michtchenko}, T.~A. 2003, \apj, 593,
  1124

\bibitem[{{Beaug{\'e}} {et~al.}(2008){Beaug{\'e}}, {Giuppone}, {Ferraz-Mello},
  \& {Michtchenko}}]{2008Beau}
{Beaug{\'e}}, C., {Giuppone}, C.~A., {Ferraz-Mello}, S., \& {Michtchenko},
  T.~A. 2008, \mnras, 385, 2151

\bibitem[{{Chambers}(1999)}]{1999C}
{Chambers}, J.~E. 1999, \mnras, 304, 793

\bibitem[{{Chambers} {et~al.}(1996){Chambers}, {Wetherill}, \& {Boss}}]{1996C}
{Chambers}, J.~E., {Wetherill}, G.~W., \& {Boss}, A.~P. 1996, Icarus, 119, 261

\bibitem[{{Chatterjee} {et~al.}(2008){Chatterjee}, {Ford}, {Matsumura}, \&
  {Rasio}}]{2008Chatterjee}
{Chatterjee}, S., {Ford}, E.~B., {Matsumura}, S., \& {Rasio}, F.~A. 2008, \apj,
  686, 580

\bibitem[{{Chauvin} {et~al.}(2004){Chauvin}, {Lagrange}, {Dumas}, {Zuckerman},
  {Mouillet}, {Song}, {Beuzit}, \& {Lowrance}}]{2004C}
{Chauvin}, G., {Lagrange}, A.-M., {Dumas}, C., {Zuckerman}, B., {Mouillet}, D.,
  {Song}, I., {Beuzit}, J.-L., \& {Lowrance}, P. 2004, \aap, 425, L29

\bibitem[{{Chiang} {et~al.}(2009){Chiang}, {Kite}, {Kalas}, {Graham}, \&
  {Clampin}}]{2009C}
{Chiang}, E., {Kite}, E., {Kalas}, P., {Graham}, J.~R., \& {Clampin}, M. 2009,
  \apj, 693, 734

\bibitem[{{Correia} {et~al.}(2005){Correia}, {Udry}, {Mayor}, {Laskar}, {Naef},
  {Pepe}, {Queloz}, \& {Santos}}]{2005C}
{Correia}, A.~C.~M., {Udry}, S., {Mayor}, M., {Laskar}, J., {Naef}, D., {Pepe},
  F., {Queloz}, D., \& {Santos}, N.~C. 2005, \aap, 440, 751

\bibitem[{{Dupuy} {et~al.}(2009){Dupuy}, {Liu}, \& {Ireland}}]{2009D}
{Dupuy}, T.~J., {Liu}, M.~C., \& {Ireland}, M.~J. 2009, \apj, 692, 729

\bibitem[{{Fukagawa} {et~al.}(2009){Fukagawa}, {Itoh}, {Tamura}, {Oasa},
  {Hayashi}, {Fujita}, {Shibai}, \& {Hayashi}}]{2009F}
{Fukagawa}, M., {Itoh}, Y., {Tamura}, M., {Oasa}, Y., {Hayashi}, S.~S.,
  {Fujita}, Y., {Shibai}, H., \& {Hayashi}, M. 2009, \apjl, 696, L1

\bibitem[{{Gayon} \& {Bois}(2008)}]{2008G}
{Gayon}, J., \& {Bois}, E. 2008, \aap, 482, 665

\bibitem[{{Gladman}(1993)}]{1993G}
{Gladman}, B. 1993, Icarus, 106, 247

\bibitem[{{Goldreich} {et~al.}(2004){Goldreich}, {Lithwick}, \& {Sari}}]{gls04}
{Goldreich}, P., {Lithwick}, Y., \& {Sari}, R. 2004, \araa, 42, 549

\bibitem[{{Go{\'z}dziewski} \& {Migaszewski}(2009)}]{2009GM}
{Go{\'z}dziewski}, K., \& {Migaszewski}, C. 2009, \mnras, 397, L16

\bibitem[{{Gray} \& {Kaye}(1999)}]{1999GK}
{Gray}, R.~O., \& {Kaye}, A.~B. 1999, \aj, 118, 2993

\bibitem[{{Hamilton} \& {Krivov}(1997)}]{1997H}
{Hamilton}, D.~P., \& {Krivov}, A.~V. 1997, Icarus, 128, 241

\bibitem[{{Hern{\'a}ndez} {et~al.}(2005){Hern{\'a}ndez}, {Calvet}, {Hartmann},
  {Brice{\~n}o}, {Sicilia-Aguilar}, \& {Berlind}}]{2005H}
{Hern{\'a}ndez}, J., {Calvet}, N., {Hartmann}, L., {Brice{\~n}o}, C.,
  {Sicilia-Aguilar}, A., \& {Berlind}, P. 2005, \aj, 129, 856

\bibitem[{{Hillenbrand} {et~al.}(1993){Hillenbrand}, {Massey}, {Strom}, \&
  {Merrill}}]{1993H}
{Hillenbrand}, L.~A., {Massey}, P., {Strom}, S.~E., \& {Merrill}, K.~M. 1993,
  \aj, 106, 1906

\bibitem[{{Iben}(1967)}]{1967I}
{Iben}, I.~J. 1967, \apj, 147, 624

\bibitem[{{Juri{\'c}} \& {Tremaine}(2008)}]{2008JT}
{Juri{\'c}}, M., \& {Tremaine}, S. 2008, \apj, 686, 603

\bibitem[{{Kalas} {et~al.}(2008){Kalas}, {Graham}, {Chiang}, {Fitzgerald},
  {Clampin}, {Kite}, {Stapelfeldt}, {Marois}, \& {Krist}}]{2008K}
{Kalas}, P., {Graham}, J.~R., {Chiang}, E., {Fitzgerald}, M.~P., {Clampin}, M.,
  {Kite}, E.~S., {Stapelfeldt}, K., {Marois}, C., \& {Krist}, J. 2008, Science,
  322, 1345

\bibitem[{{Kozai}(1962)}]{1962K}
{Kozai}, Y. 1962, \aj, 67, 591

\bibitem[{{Lafreni{\`e}re} {et~al.}(2008){Lafreni{\`e}re}, {Jayawardhana}, \&
  {van Kerkwijk}}]{2008L}
{Lafreni{\`e}re}, D., {Jayawardhana}, R., \& {van Kerkwijk}, M.~H. 2008, \apjl,
  689, L153

\bibitem[{{Lafreni{\`e}re} {et~al.}(2009){Lafreni{\`e}re}, {Marois}, {Doyon},
  \& {Barman}}]{2009L}
{Lafreni{\`e}re}, D., {Marois}, C., {Doyon}, R., \& {Barman}, T. 2009, \apjl,
  694, L148

\bibitem[{{Lagrange} {et~al.}(2009){Lagrange}, {Gratadour}, {Chauvin}, {Fusco},
  {Ehrenreich}, {Mouillet}, {Rousset}, {Rouan}, {Allard}, {Gendron}, {Charton},
  {Mugnier}, {Rabou}, {Montri}, \& {Lacombe}}]{2009Lagrange}
{Lagrange}, A., {Gratadour}, D., {Chauvin}, G., {Fusco}, T., {Ehrenreich}, D.,
  {Mouillet}, D., {Rousset}, G., {Rouan}, D., {Allard}, F., {Gendron}, {\'E}.,
  {Charton}, J., {Mugnier}, L., {Rabou}, P., {Montri}, J., \& {Lacombe}, F.
  2009, \aap, 493, L21

\bibitem[{{Lee} {et~al.}(2006){Lee}, {Butler}, {Fischer}, {Marcy}, \&
  {Vogt}}]{2006Lee}
{Lee}, M.~H., {Butler}, R.~P., {Fischer}, D.~A., {Marcy}, G.~W., \& {Vogt},
  S.~S. 2006, \apj, 641, 1178

\bibitem[{{Lee} \& {Peale}(2002)}]{2002LP}
{Lee}, M.~H., \& {Peale}, S.~J. 2002, \apj, 567, 596

\bibitem[{{Levison} {et~al.}(1998){Levison}, {Lissauer}, \& {Duncan}}]{lld98}
{Levison}, H.~F., {Lissauer}, J.~J., \& {Duncan}, M.~J. 1998, \aj, 116, 1998

\bibitem[{{Marchal} \& {Bozis}(1982)}]{1982MB}
{Marchal}, C., \& {Bozis}, G. 1982, Celestial Mechanics, 26, 311

\bibitem[{{Marcy} {et~al.}(2001){Marcy}, {Butler}, {Fischer}, {Vogt},
  {Lissauer}, \& {Rivera}}]{2001Marcy}
{Marcy}, G.~W., {Butler}, R.~P., {Fischer}, D., {Vogt}, S.~S., {Lissauer},
  J.~J., \& {Rivera}, E.~J. 2001, \apj, 556, 296

\bibitem[{{Marley} {et~al.}(2007){Marley}, {Fortney}, {Hubickyj},
  {Bodenheimer}, \& {Lissauer}}]{2007M}
{Marley}, M.~S., {Fortney}, J.~J., {Hubickyj}, O., {Bodenheimer}, P., \&
  {Lissauer}, J.~J. 2007, \apj, 655, 541

\bibitem[{{Marois} {et~al.}(2008){Marois}, {Macintosh}, {Barman}, {Zuckerman},
  {Song}, {Patience}, {Lafreni{\`e}re}, \& {Doyon}}]{2008M}
{Marois}, C., {Macintosh}, B., {Barman}, T., {Zuckerman}, B., {Song}, I.,
  {Patience}, J., {Lafreni{\`e}re}, D., \& {Doyon}, R. 2008, Science, 322, 1348

\bibitem[{{Mayor} {et~al.}(2004){Mayor}, {Udry}, {Naef}, {Pepe}, {Queloz},
  {Santos}, \& {Burnet}}]{2004Mayor}
{Mayor}, M., {Udry}, S., {Naef}, D., {Pepe}, F., {Queloz}, D., {Santos}, N.~C.,
  \& {Burnet}, M. 2004, \aap, 415, 391

\bibitem[{{Metchev} {et~al.}(2009){Metchev}, {Marois}, \& {Zuckerman}}]{2009M}
{Metchev}, S., {Marois}, C., \& {Zuckerman}, B. 2009, \apjl, 705, L204

\bibitem[{{Michtchenko} \& {Malhotra}(2004)}]{2004MM}
{Michtchenko}, T.~A., \& {Malhotra}, R. 2004, Icarus, 168, 237

\bibitem[{{Morbidelli} {et~al.}(2007){Morbidelli}, {Tsiganis}, {Crida},
  {Levison}, \& {Gomes}}]{2007Morby}
{Morbidelli}, A., {Tsiganis}, K., {Crida}, A., {Levison}, H.~F., \& {Gomes}, R.
  2007, \aj, 134, 1790

\bibitem[{{Murray} \& {Dermott}(1999)}]{1999MD}
{Murray}, C.~D., \& {Dermott}, S.~F. 1999, {Solar system dynamics} (Cambridge
  University Press)

\bibitem[{{Murray-Clay} \& {Chiang}(2006)}]{2006M}
{Murray-Clay}, R.~A., \& {Chiang}, E.~I. 2006, \apj, 651, 1194

\bibitem[{{Nesvorn{\'y}} {et~al.}(2003){Nesvorn{\'y}}, {Alvarellos}, {Dones},
  \& {Levison}}]{2003N}
{Nesvorn{\'y}}, D., {Alvarellos}, J.~L.~A., {Dones}, L., \& {Levison}, H.~F.
  2003, \aj, 126, 398

\bibitem[{{Reidemeister} {et~al.}(2009){Reidemeister}, {Krivov}, {Schmidt},
  {Fiedler}, {M{\"u}ller}, {L{\"o}hne}, \& {Neuh{\"a}user}}]{2009R}
{Reidemeister}, M., {Krivov}, A.~V., {Schmidt}, T.~O.~B., {Fiedler}, S.,
  {M{\"u}ller}, S., {L{\"o}hne}, T., \& {Neuh{\"a}user}, R. 2009, \aap, 503,
  247

\bibitem[{{Rivera} {et~al.}(2005){Rivera}, {Lissauer}, {Butler}, {Marcy},
  {Vogt}, {Fischer}, {Brown}, {Laughlin}, \& {Henry}}]{2005R}
{Rivera}, E.~J., {Lissauer}, J.~J., {Butler}, R.~P., {Marcy}, G.~W., {Vogt},
  S.~S., {Fischer}, D.~A., {Brown}, T.~M., {Laughlin}, G., \& {Henry}, G.~W.
  2005, \apj, 634, 625

\bibitem[{{Scharf} \& {Menou}(2009)}]{2009SM}
{Scharf}, C., \& {Menou}, K. 2009, \apjl, 693, L113

\bibitem[{{Su} {et~al.}(2009){Su}, {Rieke}, {Stapelfeldt}, {Malhotra},
  {Bryden}, {Smith}, {Misselt}, {Moro-Martin}, \& {Williams}}]{srs+09}
{Su}, K.~Y.~L., {Rieke}, G.~H., {Stapelfeldt}, K.~R., {Malhotra}, R., {Bryden},
  G., {Smith}, P.~S., {Misselt}, K.~A., {Moro-Martin}, A., \& {Williams}, J.~P.
  2009, \apj, 705, 314

\bibitem[{{Terquem} \& {Papaloizou}(2007)}]{2007T}
{Terquem}, C., \& {Papaloizou}, J.~C.~B. 2007, \apj, 654, 1110

\bibitem[{{van Leeuwen}(2007)}]{2007VL}
{van Leeuwen}, F. 2007, \aap, 474, 653

\bibitem[{{Veras} {et~al.}(2009){Veras}, {Crepp}, \& {Ford}}]{2009VCF}
{Veras}, D., {Crepp}, J.~R., \& {Ford}, E.~B. 2009, \apj, 696, 1600

\bibitem[{{Vogt} {et~al.}(2005){Vogt}, {Butler}, {Marcy}, {Fischer}, {Henry},
  {Laughlin}, {Wright}, \& {Johnson}}]{2005V}
{Vogt}, S.~S., {Butler}, R.~P., {Marcy}, G.~W., {Fischer}, D.~A., {Henry},
  G.~W., {Laughlin}, G., {Wright}, J.~T., \& {Johnson}, J.~A. 2005, \apj, 632,
  638

\bibitem[{{Williams} \& {Andrews}(2006)}]{2006WA}
{Williams}, J.~P., \& {Andrews}, S.~M. 2006, \apj, 653, 1480

\bibitem[{{Zhou} {et~al.}(2007){Zhou}, {Lin}, \& {Sun}}]{2007Z}
{Zhou}, J.-L., {Lin}, D.~N.~C., \& {Sun}, Y.-S. 2007, \apj, 666, 423

\end{thebibliography}

\clearpage

\begin{deluxetable}{ccccccccc}
\tablecaption{Log-book of integrations: I. Orbital Elements\label{tab:logI}}
\tabletypesize{\normalsize}
\tablewidth{0pt}

\tablehead{
\colhead{Integration} &
\colhead{$M_\star$ ($M_\odot$)} &
\colhead{$M_p$ ($M_\odot$)} &
\colhead{$a$ (AU)} &
\colhead{$e$} &
\colhead{$i$} &
\colhead{$\omega$} &
\colhead{$\Omega$} &
\colhead{$M$*} 
}

\startdata  
\S\ref{sec:astrom}, D &$1.86$ & $0.0067$ & $81.04$ & $0.0$ & $0^\circ$ & $0^\circ$ & $0^\circ$ & $61.30^\circ$ \\
Fig. \ref{fig:185}   & & $0.0095$ & $38.16$ & $0.0$ & $0^\circ$ & $0^\circ$ & $0^\circ$ & $315.29^\circ$ \\
           & & $0.0095$ & $27.69$ & $0.0$ & $0^\circ$ & $0^\circ$ & $0^\circ$ & $204.92^\circ$ \\
\hline
\S\ref{sec:mmr} &$1.5$ & $0.0095$ & $37.97$ & $0.0$ & $0^\circ$ & $0^\circ$ & $0^\circ$ & $226.99^\circ$ \\
two-planet resonance & & $0.0095$ & $23.32$ & $0.09$ & $0^\circ$ & $344.0^\circ$ & $0^\circ$ & $126.00^\circ$ \\

\hline
\S\ref{sec:mmr} &$1.5$ & $0.0067$ & $67.91$ & $0.0$ & $0^\circ$ & $0^\circ$ & $0^\circ$ & $332.36^\circ$ \\
Fig. \ref{fig:unstabres} & & $0.0095$ & $37.97$ & $0.0$ & $0^\circ$ & $0^\circ$ & $0^\circ$ & $226.99^\circ$ \\
           & & $0.0095$ & $23.32$ & $0.09$ & $0^\circ$ & $344.0^\circ$ & $0^\circ$ & $126.00^\circ$\\
\hline
\S\ref{sec:mmr} &$1.5$ & $0.0067$ & $67.91$ & $0.0$ & $0^\circ$ & $0^\circ$ & $0^\circ$ & $332.36^\circ$ \\
Fig. \ref{fig:boundres} & & $0.0095$ & $37.97$ & $0.0$ & $0^\circ$ & $0^\circ$ & $0^\circ$ & $226.99^\circ$ \\
           & & $0.0095$ & $23.42$ & $0.09$ & $0^\circ$ & $344.0^\circ$ & $0^\circ$ & $126.00^\circ$\\
\hline
\S\ref{sec:mmr}** & $1.5$ & $0.0067x$ & $67.91$ & $0.0$ & $0^\circ$ & $0^\circ$ & $0^\circ$ & $332.36^\circ$ \\
Fig. \ref{fig:massfactor}    & & $0.0095x$ & $37.97$ & $0.0$ & $0^\circ$ & $0^\circ$ & $0^\circ$ & $226.99^\circ$ \\
non-resonant           & & $0.0095x$ & $24.44$ & $0.0$ & $0^\circ$ & $344.0^\circ$ & $0^\circ$ & $126.00^\circ$ \\
           &  &              & $\pm10^{-4}$ & $\pm10^{-4}$ & $\pm0.01^\circ$ & $\pm0.01^\circ$ & $\pm0.01^\circ$ & $\pm0.01^\circ$ \\   
\hline
\S\ref{sec:mmr}** & $1.5$ & $0.0067x$ & $67.91$ & $0.0$ & $0^\circ$ & $0^\circ$ & $0^\circ$ & $332.36^\circ$ \\
Fig. \ref{fig:massfactor}    & & $0.0095x$ & $37.97$ & $0.0$ & $0^\circ$ & $0^\circ$ & $0^\circ$ & $226.99^\circ$ \\
single resonance           & & $0.0095x$ & $23.32$ & $0.09$ & $0^\circ$ & $344.0^\circ$ & $0^\circ$ & $126.00^\circ$ \\
           &  &              & $\pm10^{-4}$ & $\pm10^{-4}$ & $\pm0.01^\circ$ & $\pm0.01^\circ$ & $\pm0.01^\circ$ & $\pm0.01^\circ$ \\   
\hline
\S\ref{sec:mmr}** & $1.5$ & $0.0067x$ & $67.91$ & $0.002$ & $0^\circ$ & $180^\circ$ & $0^\circ$ & $180^\circ$ \\
Fig. \ref{fig:massfactor}    & & $0.0095x$ & $37.97$ & $0.005$ & $0^\circ$ & $0^\circ$ & $0^\circ$ & $0^\circ$ \\
double resonance           & & $0.0095x$ & $23.52$ & $0.083$ & $0^\circ$ & $180^\circ$ & $0^\circ$ & $0^\circ$ \\
           &  &              & $\pm10^{-4}$ & $\pm10^{-4}$ & $\pm0.01^\circ$ & $\pm0.01^\circ$ & $\pm0.01^\circ$ & $\pm0.01^\circ$ \\   
\hline
\S\ref{sec:nonplanar}*** &$1.5$ & $0.0048$ & $67.91$ & $0.0$ & $i$ & $0^\circ$ & $\Omega_b$ & $62.36^\circ$ \\
Fig. \ref{fig:mutinc}     & & $0.0067$ & $37.97$ & $0.0$ & $0^\circ$ & $0^\circ$ & $0^\circ$ & $316.99^\circ$ \\
           & & $0.0067$ & $24.44$ & $0.0$ & $i$ & $0^\circ$ & $\Omega_d$ & $200.23^\circ$ \\
\enddata
\vspace{0.1 in}
\tablecomments{ For each integration or suite of integrations listed, the three lines specify the initial conditions for planets b, c, and d respectively.  These are the values we input to the integrator {\slshape Mercury} \citep{1999C}.  \\(*) Mean anomaly. \\ (**) In this series of integrations the planetary masses were scaled by various factors ($x$).  To generate Figure~\ref{fig:massfactor}, small, random (Gaussian) components with the indicated standard deviation were added to the orbital elements for a statistical sample of the chaotic outcomes. \\ (***) This series of integrations spanned a grid of three orientation parameters, $i$ (common to planets b and d), $\Omega_b$, and $\Omega_d$.  The values were $i \in [0^\circ, 2^\circ, 4^\circ,..., 20^\circ, 30^\circ, 40^\circ, ..., 180^\circ]$; $\Omega_b$ and $\Omega_d \in [0^\circ, 60^\circ, 120^\circ,..., 300^\circ]$.
\vspace{0.2 in}
}
\end{deluxetable}

\clearpage           

\begin{deluxetable}{ccccccccc}
\tablecaption{Log-book of integrations: II. State Vectors\label{tab:logII}}
\tabletypesize{\normalsize}
\tablewidth{0pt}

\tablehead{
\colhead{Integration} & 
\colhead{$M_\star$} & 
\colhead{$M_p$} & 
\colhead{$x_N$} & 
\colhead{$x_E$} & 
\colhead{$x_Z$} & 
\colhead{$v_N$} & 
\colhead{$v_E$} & 
\colhead{$v_Z$}  \\
\colhead{ } & 
 \multicolumn{2}{c}{($M_\odot$)} &
 \multicolumn{3}{c}{(AU)} &
 \multicolumn{3}{c}{(10$^{-3}$ AU day$^{-1}$)} 
}

\startdata  
\S\ref{sec:intro}, nominal & $1.5$ & $0.0067$ & $60.16$&$31.50$&$0.0$&$1.186$&$-2.265$&$0.0$ \\
\S\ref{sec:astrom}, A      & & $0.0095$ & $-25.90$ & $27.76$ & $0.0$ & $2.500$ & $2.332$ & $0.0$ \\
Fig. \ref{fig:ae},\ref{fig:data}    & & $0.0095$ & $-8.45$ & $-22.93$ & $0.0$ & $-3.998$ & $1.473$ & $0.0$  \\
\hline
\S\ref{sec:masses} & $1.5$ & $0.0067x$ & $60.16$&$31.50$&$0.0$&$1.186$&$-2.265$&$0.0$ \\
Fig. \ref{fig:massmult}*  & & $0.0095x$ & $-25.90$ & $27.76$ & $0.0$ & $2.496$ & $2.329$ & $0.0$ \\
           & & $0.0095x$ & $-8.45$ & $-22.93$ & $0.0$ & $-3.996$ & $1.473$ & $0.0$  \\
\hline
\S\ref{sec:stable}** & $1.5$ & $0.00435$ & $60.16$&$31.50$&$0.0$&$1.186$&$-2.265$&$0.0$ \\
Fig. \ref{fig:instabgridelm},\ref{fig:protect}   & & $0.0062$ & $-25.90$ & $27.76$ & $0.0$ & $2.496$ & $2.329$ & $0.0$ \\
           & & $0.0062$ & $-8.45$ & $-22.93$ & $0.0$ & $-3.996 \gamma$ & $1.473 \gamma$ & $0.0$  \\
\hline
\S\ref{sec:mode}& $1.60905$ & $0.0048$ & $60.18$ & $31.47$ & $27.21$ & $1.464$ & $-2.244$ & $0.4295$ \\
Fig. \ref{fig:seclock}  &           & $0.0067$ & $-25.94$ & $27.89$ & $-8.527$ & $ 2.398$ & $2.460$ & $1.178$ \\
        &           & $0.0067$ &$-8.601$ & $-22.90$ & $-5.311$ & $-3.577$ & $1.891$ & $-1.327$\\
\hline
\S\ref{sec:arbitrary} & $1.54796$ & $0.0048$ & $57.54$ & $30.22$ & $9.765$ & $1.293$ & $-2.034$ & $0.2204$ \\ 
Fig. \ref{fig:noncoplanar}       &           & $0.0067$ & $-24.74$ & $26.61$ & $-23.15$ & $2.228$ & $2.327$ & $1.334$ \\
        &           & $0.0067$ & $-8.084$ & $-22.00$ & $-9.609$ & $-3.490$ & $1.518$ & $0.2634$ 
\enddata
\vspace{0.1 in}
\tablecomments{ (*) In this series of integrations the planetary masses were scaled by various factors ($x$).  Some reported integrations had only two planets: they were missing either the first planet listed (b) or the last planet listed (d).  \\ (**) In this series of integrations the innermost planet had a non-zero eccentricity.  This was implemented by multiplying its initial velocity from the face-on, coplanar model by $\gamma \equiv \sqrt{2-24.44 \rm{AU}/a_d}$.  Two series were performed, one with all three planets and one without the first planet listed (b).
\vspace{0.2 in}
}
\end{deluxetable}

\end{document}